\documentclass[11pt,letterpaper]{article}
\pdfoutput=1
\usepackage{jheppub} 
\usepackage{mathrsfs}
\usepackage{slashed}
\usepackage{caption}
\usepackage{epstopdf}
\usepackage[normalem]{ulem}
\usepackage[bottom]{footmisc}
\usepackage{subcaption}
\usepackage{bbold}
\usepackage{titlesec}
\usepackage{threeparttable}
\usepackage{booktabs}
\usepackage{changepage}
\usepackage[utf8]{inputenc}
\usepackage{dsfont}
\usepackage{grffile}
\usepackage{graphicx}  % needed for figures
\usepackage{dcolumn}   % needed for some tables
\usepackage{bm}        % for math
\usepackage{amssymb}   % for math
\usepackage{setspace}
\usepackage{amsmath, amssymb, setspace}
\usepackage{array}
\usepackage{booktabs}
\usepackage{caption}
\usepackage{indentfirst}
\usepackage{float}
\usepackage{lmodern}
\usepackage{multirow}
\usepackage{soul}
\usepackage[normalem]{ulem}
\usepackage{braket}
\usepackage{comment}
\usepackage{appendix}  
\usepackage{feynmp-auto}
\graphicspath{{figs/}}

\newcommand{\eg}{{\it e.g.}}

\title{\huge{Hunting Nonstandard Neutrino Interactions and Leptoquarks in Dark Matter Experiments}} 

\author[a]{Thomas Schwemberger,}
\author[b,c,d,e]{Volodymyr Takhistov,}
\author[a,f]{Tien-Tien Yu}

\affiliation[a]{Department of Physics and Institute for Fundamental Science, University of Oregon\\
Eugene, Oregon 97403, USA
}
\affiliation[b]{International Center for Quantum-field Measurement Systems for Studies of the Universe and Particles (QUP, WPI),
High Energy Accelerator Research Organization (KEK), \\ Oho 1-1, Tsukuba, Ibaraki 305-0801, Japan}
\affiliation[c]{Theory Center, Institute of Particle and Nuclear Studies (IPNS), High Energy Accelerator Research Organization (KEK), Tsukuba 305-0801, Japan}
\affiliation[d]{Kavli Institute for the Physics and Mathematics of the Universe (WPI), \\ The University of Tokyo Institutes for Advanced Study, The University of Tokyo, \\ Kashiwa, Chiba 277-8583, Japan}
\affiliation[e]{Graduate University for Advanced Studies (SOKENDAI),  
1-1 Oho, Tsukuba, Ibaraki 305-0801, Japan} 
\affiliation[f]{Center for Cosmology and Particle Physics, Department of Physics, New York University, New York, NY 10003, USA}

\emailAdd{tschwem2@uoregon.edu}
\emailAdd{vtakhist@post.kek.jp}
\emailAdd{tientien@uoregon.edu} 

\abstract{
Next generation direct dark matter (DM) detection experiments will have unprecedented capabilities to explore coherent neutrino-nucleus scattering (CE$\nu$NS) complementary to dedicated neutrino experiments.
We demonstrate that future DM experiments can effectively probe nonstandard neutrino interactions (NSI) mediated by scalar fields in the scattering of solar and atmospheric neutrinos. We set first limits on $S_1$ leptoquark models that result in sizable $\mu-d$ and $\tau-d$ sector neutrino NSI CE$\nu$NS contributions using LUX-ZEPLIN (LZ) data. As we show, near future DM experiments reaching $\sim \mathcal{O}(100)$ton-year exposure, such as argon-based ARGO and xenon-based DARWIN, can probe parameter space of leptoquarks beyond the reach of current and planned collider facilities. We also analyze for the first time prospects for testing NSI in lead-based detectors.
We discuss the ability of leptoquarks in the parameter space of interest to also explain the neutrino masses and $(g-2)_\mu$ observations.
}

\begin{document}

\preprint{KEK-QUP-2023-0007, KEK-TH-2516, KEK-Cosmo-0309, IPMU23-0009}
 \maketitle
\flushbottom
 
\section{Introduction}
Dark matter (DM) constitutes $\sim85\%$ of all matter in the Universe and understanding its nature is one of the most important goals of science. Significant efforts have been devoted over the past several decades to detect the scattering of particle DM from the Galactic DM halo with target nuclei in underground laboratories. Historically, the focus has been on probing interactions of weakly-interacting massive particles (WIMPs) with masses $\sim 100$~GeV-TeV, using large ton-scale experiments based on liquid xenon, \eg~LUX-ZEPLIN (LZ)~\cite{LZ:2022ufs}, XENONnT~\cite{XENON:2023sxq}, and liquid argon, \eg~DEAP-3600~\cite{DEAP:2019yzn} and DarkSide-50~\cite{DarkSide-50:2022qzh}, detectors. Not only are these detectors very sensitive to the scattering of DM particles with Standard Model (SM) targets, they can also be sensitive to and provide new insights into additional phenomena both within and beyond the SM (BSM). 

As direct DM detection experiments continue to improve their sensitivity, they will inevitably encounter an irreducible background stemming from coherent elastic neutrino-nucleus scattering (CE$\nu$NS).
Predicted decades ago~\cite{Freedman:1973yd,Drukier:1984vhf}, CE$\nu$NS interactions have been directly observed by COHERENT in 2017 using neutrinos generated by pions that decay at rest~\cite{COHERENT:2017ipa}.\footnote{Efforts to observe CE$\nu$NS using reactor anti-neutrinos are also being pursued~\cite{CONUS:2021dwh,CONNIE:2016nav,NUCLEUS:2019igx,Billard:2016giu}.}
Interactions of neutrinos originating from natural sources, such as solar neutrinos from the Sun and atmospheric neutrinos from cosmic ray collisions with the atmosphere, are 
indistinguishable on an event-by-event basis from DM-induced scattering events depositing energy within detectors~(\eg ~\cite{Billard:2014, Essig:2018tss,Gelmini:2018ogy}). 
The degeneracy between neutrino and DM scattering results in the so-called ``neutrino fog" (formerly known as the ``neutrino floor").
Analyses of event statistics and temporal modulation, such as with sidereal day, can help disentangle the degeneracy.

Since the discovery of neutrino oscillations~\cite{Super-Kamiokande:1998kpq}, a multitude of experiments have confirmed and further developed the SM picture of neutrino interactions and 3-flavor oscillations~\cite{ParticleDataGroup:2022pth}.
With neutrino physics entering the precision era in a broad, world-wide program including a variety of upcoming major next generation experiments such as DUNE and Hyper-Kamiokande, significant attention is being devoted to probing new neutrino physics.
Sizable novel ``non-standard'' neutrino interactions (NSI) can readily appear~(see \eg~\cite{Proceedings:2019qno} for review) in broad classes of motivated models, such as those based on SM extensions by an additional $U(1)$ gauge symmetry and new mediators (\eg~\cite{Farzan:2016wym}). 

Along with precision tests of SM parameters, observations of CE$\nu$NS, such as data from COHERENT, allow for a unique probe of neutrino NSI and related theoretical models~\cite{Barranco:2005yy,Scholberg:2005qs,Miranda:2020tif,Liao:2017uzy,Giunti:2019xpr,Khan:2021wzy,Denton:2020hop,Farzan:2018gtr,Arcadi:2019uif,Billard:2018jnl,Bertuzzo:2021opb,delaVega:2021wpx,Banerjee:2021laz,Cadeddu:2020nbr,Flores:2020lji,Denton:2018xmq}.  
Upcoming next generation DM experiments will have sensitivity to CE$\nu$NS that allows for unique opportunities to test novel neutrino interactions in complementary ways to dedicated neutrino experiments. Studies already initiated exploration of the reach of DM detectors to neutrino NSI~\cite{Harnik:2012ni,Cerdeno:2016sfi} mediated by novel scalar, pseudoscalar, vector, and axial-vector particles. 
Limits on solar neutrino NSI from observations of previous DM experiments have also been explored,\footnote{Neutrino NSI can also be readily probed through electron scattering in DM experiments, which has been recently exploited in attempt to address excess in the XENON1T experiment data~\cite{Boehm:2020ltd}. However, in this work we focus solely on CE$\nu$NS interactions.} such as from LUX experiment data~\cite{Bertuzzo:2017tuf}.  

In this work, we expand on previous studies and demonstrate the power of DM experiments to probe neutrino NSI. In the case of heavy scalar mediators, the effective interaction becomes a dimension six operator which can either be the familiar NSI modeled after the SM vector-interaction ($\bar{\nu}\gamma^\alpha P_L \nu \bar{q}\gamma_\alpha P_L q$), which is a vector NSI, or a scalar NSI ($\bar{\nu}\nu \bar{q}q$). Our analysis advances the results in the literature in several important ways.
While SM interactions of atmospheric neutrinos have been explored in DM experiments~\cite{Newstead:2020fie}, we study for the first time their potential NSI effects.
Furthermore, we explore first constraints on leptoquarks (LQs) and associated theories that can lead to sizable solar and atmospheric neutrino NSI effects in DM experiments, and for the first time use LZ data to set constraints on LQs. We discuss the complementarity of LQ searches in DM experiments and other venues, especially colliders, and connect them with models aiming to address flavor physics observations~\cite{angelescu_closing_2018, Lee:2021jdr, FileviezPerez:2021lkq}, $(g-2)_\mu$, and neutrino mass generation \cite{Dorsner:2017wwn, zhang_radiative_2021, PhysRevD.LQ3}.
Our analyses are complementary to studies of NSI in DM experiments associated with light mediators, such as~Ref.~\cite{LI2022115737, PhysRevD.106.013001,Harnik:2012ni,deNiverville:2015mwa,Cerdeno:2016sfi} or ~Ref.~\cite{Schwemberger:2022fjl}, the latter of which demonstrated that low-threshold ($\sim$ eV) DM detectors are sensitive to light ($\sim$ keV - MeV) mediators where the low energy limit provides an enhancement of the recoil rate.  

The paper is organized as follows. In the Sec. \ref{sec:flux}, we describe the neutrino flux, its flavor composition, and its signal in DM detectors. In Sec. \ref{sec:BSM} we show the effects of neutrino NSIs and the effective theories that produce such interactions. In Sec. \ref{sec:LQ} we discuss the details of a UV completion of the effective models as LQs including effects on neutrino masses, $(g-2)_\mu$, and neutrino oscillations. In Sec. \ref{sec:disc_reach} we demonstrate the ability of DM detectors to observe the effective models described in Sec. \ref{sec:BSM}, calculate constraints from the LZ experiment, and project the discovery reach of future DM detectors based on xenon, argon, and lead.

\section{Standard Model Neutrino Scattering}\label{sec:flux}

\subsection{Interaction rates and CE\texorpdfstring{$\nu$}{v}NS}
\label{sec:rates}

In DM experiments, the recoil rate of neutrinos as a function of the nuclear recoil energy depends on 
both the physics of the scattering process encapsulated in the cross-section, neutrino energy and flux, as well as the properties of the detector such as target material, size, efficiency, energy threshold, and energy resolution. The general differential recoil rate is given by
\begin{equation}\label{eq:gen_scattering}
    \frac{dR}{dE'_R} = N_T \int_{E_\nu^{\rm min}} \frac{d\sigma}{dE'_R} \frac{dN_\nu}{dE_\nu}dE_\nu~,
\end{equation}
where $dN_\nu/dE_\nu$ is the neutrino flux, $N_T$ is the number of target nuclei per unit mass (ton) of detector material, $d\sigma/d{E'}_R$ is process cross-section and for SM CE$\nu$NS it is given by Eq.~\eqref{eq:SM_cross}. Here,
primed energies represent the scattering energy of the interaction while the unprimed energies are those actually measured by the detector. The effects of new physics on the interactions will be explored in Sec.~\ref{sec:BSM}.
From kinematics, the minimum required neutrino energy to induce a recoil energy $E_R^{\prime}$ of a nucleus with mass $m_N$ is 
\begin{equation}
\label{eq:emin}
E_\nu^{\rm min} =  \dfrac{1}{2}\left({E'}_R+\sqrt{{E'}_R^2 + 2{E'}_R m_N}\right)~.
\end{equation} 

We incorporate the detection efficiency and energy resolution of the detector by integrating over $E'_R$ convoluted with the efficiency and a Gaussian function representing uncertainty
\begin{equation}
    \frac{dR}{dE_R} = \int_0^\infty \frac{dR}{dE'_R} \frac{\eta(E'_R)}{\zeta(E'_R)\sqrt{2\pi}}e^{-\frac{(E_R - E'_R)^2}{2\zeta^2(E'_R)}}dE'_R\, ,
\end{equation}
where for $\eta(E'_R)$ we will consider data-driven efficiency function of Eq.~\eqref{eq:efficiency} and $\zeta(E'_R)$ is the energy resolution that we conservatively assume to be $\mathcal{O}(10)\%$ for all energies of interest.\footnote{Our analysis can be readily adapted for other efficiency and resolution functions.}

\begin{figure}
    \centering
    \includegraphics[width=0.6\textwidth]{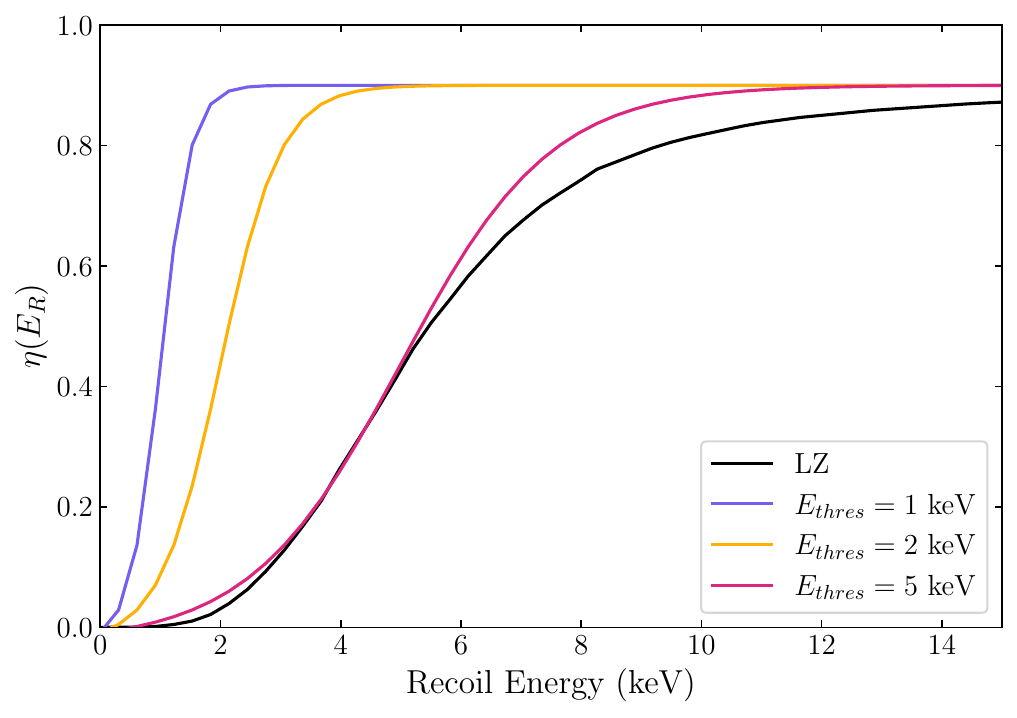}
    \caption{Efficiency models for 1, 2, and 5 keV thresholds compared to LZ experiment data analysis~\cite{LZ:2022ufs}. The sensitivity of DM detectors to NSI is strongly dependent on their energy threshold (\eg~\cite{Schwemberger:2022fjl}). We consider the 2 keV threshold model to be an optimistic efficiency for such near future detectors.}
    \label{fig:expeff}
\end{figure}

Focusing on the data analysis from LZ experiment as a characteristic target~\cite{LZ:2022ufs}, we model the detection efficiency with a hyperbolic tangent as
\begin{equation}\label{eq:efficiency}
    \eta(E) = \frac{\eta_0}{1+\zeta}\left(\tanh{\xi\left(\frac{E}{E_{\rm th}}-1\right)}+\zeta\right)~,
\end{equation} 
where $\eta_0$ is the maximum efficiency at high energy, $\xi$ sets the rate of the efficiency rise, $E_{\rm th}$ is the threshold energy, and $\zeta$ enforces that the efficiency vanishes at zero recoil energy. We fix $\xi = 2$, which requires $\zeta \approx 0.964$. To approximately match the LZ efficiency, we use $\eta_0 = 0.9$ such that the only remaining parameter is $E_{\rm th}$, which we vary between 1 and 5 keV. We display the efficiency for a selection of threshold energies in Fig.~\ref{fig:expeff}. 

Neutrinos that do not possess sufficient energy to discriminate individual nuclear constituents interact with the whole target nucleus via CE$\nu$NS mediated by the weak neutral current~\cite{Freedman:1973yd,Drukier:1984vhf}. For low momentum transfer $|q|$ satisfying $1/|q| \gg R$, where $R$ is the nuclear radius, the target nucleons interact in phase. CE$\nu$NS is particularly relevant for neutrino sources with $E_{\nu} < 50$~MeV, above which other interactions become significant (see \eg~\cite{Formaggio:2012cpf} for review). Within the SM, CE$\nu$NS interactions are given by
\begin{equation}\label{eq:SM_cross}
    \left.\frac{d\sigma_N}{dE_R}\right\vert_{SM} \hspace{0.5mm} = \dfrac{G_F^2}{4\pi}F^2(E_R) Q_v^2 m_N\left(1-\frac{m_N E_R}{2E_\nu^2}\right)\, ,
\end{equation}
where $G_F$ is the Fermi constant, $Q_v = N - Z(1-4s^2_w)$ is the
weak nuclear charge, $N$ is the number of neutrons, $Z$ is the number of protons, and $s_w$ is the sine of the Weinberg angle. 
For the atomic mass number $A = N + Z$ and $m_n = 931$~MeV, the nucleus mass is $m_N \approx A m_n$. 
Here, we employ the Helm nuclear form factor \cite{LEWIN199687, PhysRev.104.1466}
\begin{equation}
F^2(E_R) = \Big[\dfrac{3 j_1(q R_{\rm eff})}{q R_{\rm eff}}\Big]^2 e^{-(q s)^2}~,
\end{equation} 
where $j_1$ denotes spherical Bessel function, $q = \sqrt{2 m_N E_R}$ is the transferred 3-momentum, $R_{\rm eff} =\sqrt{(1.23A^{1/3} - 0.6)^2 + 7\pi^2r_0^2/3 - 5s^2}$ is the effective nuclear radius, $s=0.9$ fm is the skin thickness and $r_0=0.52$ fm is fit numerically from muon scattering data.
With $s_w^2 = 0.232$~\cite{ParticleDataGroup:2022pth} CE$\nu$NS interactions scale as $\sim N^2$, signifying sizable enhancement of interaction rates for heavy target materials such xenon, argon, or lead. 

\subsection{Neutrino sources}

A variety of neutrino sources can contribute to irreducible background ({\it i.e.} giving rise to ``neutrino floor/fog'')  in direct DM detection experiments, including solar, reactor, geo-,
diffusive supernovae background (DSNB) as well as atmospheric neutrinos. 
For reference, we consider benchmark detector located at SNOLab in Sudbury, Canada that is among likely sites to host next generation DM experiments.\footnote{The SNOLab experimental site is located at 46$^{\circ}$28'19'' N, 81$^{\circ}$11'12'' W with a 6010 meter water equivalent depth.}

\begin{figure}[t]
    \centering
    \includegraphics[width=0.65\linewidth]{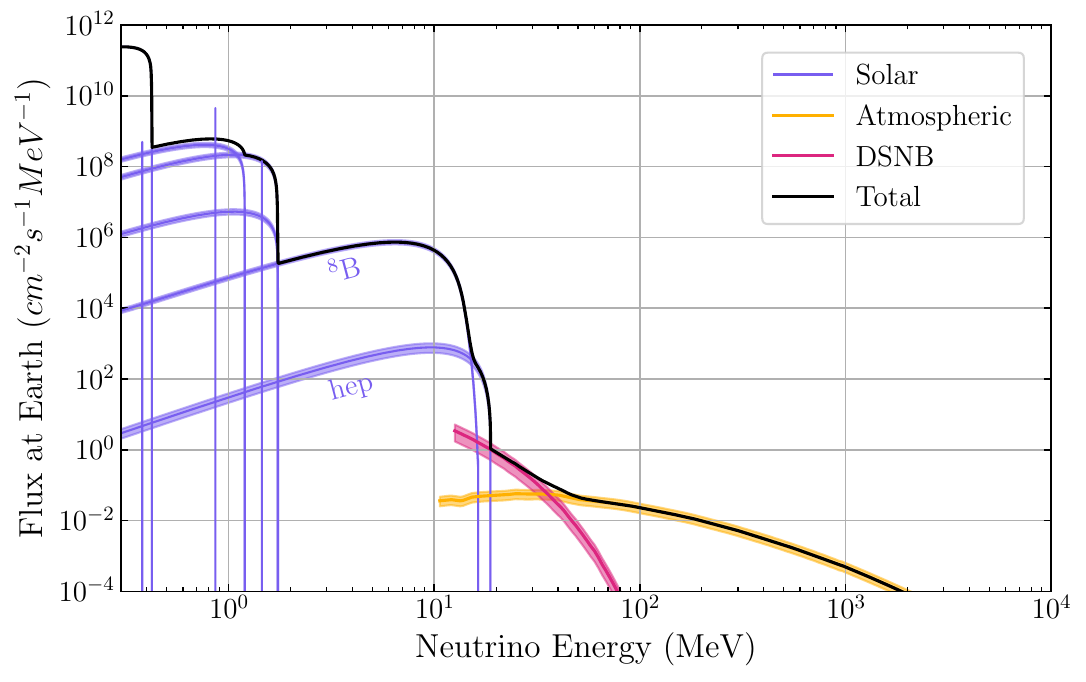}
    \caption{Neutrino flux contributions including solar, atmospheric, and DSNB components and their respective uncertainties on the surface of the Earth considering SNOLab detector location. See text and Table~\ref{tab:nu_flux} for additional details.}
    \label{fig:flux}
\end{figure}

The integrated fluxes for various neutrino components and their uncertainties (rounded to percent level) are listed in Table \ref{tab:nu_flux}. Contributions from atmospheric, reactor and geo-neutrinos will depend on the location of the experiment.
Depending on the experimental energy threshold for nuclear recoil energy $E_r$, only fluxes resulting in neutrinos with maximum energy $E_{\nu}^{\rm max} > E_{\nu}^{\rm min}$, where $E_{\nu}^{\rm min}$ is given by Eq.~\eqref{eq:emin}, will contribute. In Table~\ref{tab:nu_flux} we display the minimum required energy thresholds in xenon-based and argon-based DM experiments\footnote{Considering dominant isotopes, we take $A(Z)$ = 130 (54) for xenon and $A(Z)$ = 40(18) for argon.} to detect a particular neutrino source flux. 
We display the combined solar, DSNB and atmospheric neutrino flux contributions in Fig.~\ref{fig:flux}, along with shaded bands depicting their respective uncertainties. 

The keV-level threshold recoils in representative xenon-based experiment require that $E_{\nu}^{\rm min} \gtrsim \mathcal{O}$(few$\times$MeV), such that solar (hep and $^8$B) background contributions dominate in the region of interest for non-standard neutrino interactions. We include the subdominant atmospheric neutrinos as they are useful in specific UV complete NSI models. A representative figure for scalar-mediated non-standard neutrino interactions are displayed in Fig.~\ref{fig:recoil_rates}, further discussed below in Sec.~\ref{sec:BSM}.

 \begin{table*}[tbp]
\setlength{\extrarowheight}{2pt}
  \setlength{\tabcolsep}{10pt}
  \begin{center}
  \begin{threeparttable}
	\begin{tabular}{|l|l|c|c|l|}  \hline
	~~~~~~Neutrino & ~~~~~~~Total Flux & $E_{\nu}^{\rm max}$ & $E_{\rm th}^{\rm Xe}$ ($E_{\rm th}^{\rm Ar}$) & ~~~~Reference \\ 
    ~~~~Contribution & ~~~~~~~[cm$^{-2}$ s$^{-1}$]               & [MeV] & [keV] & ~~~~~(model) \\ \hline
	\hline
    Solar ($\nu_e$, pp)       & $5.98 (1 \pm 0.01)  \times 10^{10}$&   $0.42$  & $2.99(9.81)\times 10^{-3}$ & \cite{Vinyoles:2016djt} (B16-GS98) \\ \hline
	Solar ($\nu_e$, pep)      & $1.44 (1 \pm  0.01) \times 10^{8}$ &   $1.45$  & $3.42(11.2)\times 10^{-2}$ & \cite{Vinyoles:2016djt} (B16-GS98)\\ \hline
    Solar ($\nu_e$, hep)      & $7.98 (1 \pm  0.30) \times 10^{3}$ &   $18.77$ & $5.74(18.9)$               & \cite{Vinyoles:2016djt} (B16-GS98) \\ \hline
    Solar ($\nu_e$, $^{7}$Be) & $4.93 (1 \pm  0.06) \times 10^{9}$ &   $0.39$  & $1.20(3.96)\times 10^{-2}$ & \cite{Vinyoles:2016djt} (B16-GS98) \\ \hline
    Solar ($\nu_e$, $^{8}$B)  & $5.46 (1 \pm  0.12) \times 10^{6}$ &   $16.80$ & $4.58(15.1)$               & \cite{Vinyoles:2016djt} (B16-GS98) \\ \hline
    Solar ($\nu_e$, $^{13}$N) & $2.78 (1 \pm  0.15) \times 10^{8}$ &   $1.20$  & $2.33(7.65)\times 10^{-2}$ & \cite{Vinyoles:2016djt} (B16-GS98) \\ \hline
    Solar ($\nu_e$, $^{15}$O) & $2.05 (1 \pm  0.17) \times 10^{8}$ &   $1.73$  & $0.0486(0.160)$            & \cite{Vinyoles:2016djt} (B16-GS98) \\ \hline
    Solar ($\nu_e$, $^{17}$F) & $5.29 (1 \pm  0.20) \times 10^{6}$ &   $1.74$  & $0.0492(0.161)$            & \cite{Vinyoles:2016djt} (B16-GS98) \\ \hline
    \hline
    Atm. ($\nu_e$)\tnote{1}   & $1.27 (1 \pm  0.50)\times 10^{1}$ & $944$    & $1.47(5.00)\times 10^4$     & \cite{Battistoni:2005pd} (FLUKA) \\ \hline   
    Atm. ($\overline{\nu}_e$) & $1.17 (1 \pm  0.50)\times 10^{1}$ & $944$    & $1.47(5.00)\times 10^4$     & \cite{Battistoni:2005pd} (FLUKA) \\ \hline   
    Atm. ($\nu_{\mu}$)        & $2.46 (1 \pm  0.50)\times 10^{1}$ & $944$    & $1.47(5.00)\times 10^4$     & \cite{Battistoni:2005pd} (FLUKA) \\ \hline   
    Atm. ($\overline{\nu}_{\mu}$) & $2.45 (1 \pm  0.50)\times 10^{1}$ & $944$ & $1.47(5.00)\times 10^4$ & \cite{Battistoni:2005pd} (FLUKA) \\ \hline
    \hline
DSNB ($\nu_e$)\tnote{3}      & $4.55 (1 \pm  0.50)\times 10^{1}$ & $36.90$ & $22.1(72.7)$ & \cite{Horiuchi:2008jz} (th.~avrg.)\tnote{2} \\ \hline
DSNB ($\overline{\nu}_e$)    & $2.73 (1 \pm  0.50)\times 10^{1}$ & $57.01$ & $52.8(174)$  & \cite{Horiuchi:2008jz} (th.~avrg.)\tnote{2} \\ \hline
DSNB ($\nu_x$)\tnote{4}      & $1.75 (1 \pm  0.50)\times 10^{1}$ & $81.91$ & $109(359)$   & \cite{Horiuchi:2008jz} (th.~avrg.)\tnote{2} \\ \hline
\hline
Reactor ($\overline{\nu}_e$, $^{235}$U)\tnote{5} & $1.88(1 \pm 0.08) \times 10^5$  & 10.00  & $1.62(5.33)$ & \cite{Gelmini:2018ogy}(combined)\tnote{6} \\
\hline
\hline
Geo. ($\overline{\nu}_e$, $^{40}$K)   & $2.19 (1 \pm 0.17) \times 10^{7}$ &  1.32  & $2.83(9.29)\times 10^{-2}$ & \cite{huang:2013geomodel} (global)\tnote{6} \\
\hline
Geo. ($\overline{\nu}_e$, $^{238}$U)  & $4.90 (1 \pm 0.20) \times 10^{6}$ &  3.99  & $0.259(0.849)$             & \cite{huang:2013geomodel} (global)\tnote{6} \\
\hline
Geo. ($\overline{\nu}_e$, $^{232}$Th) & $4.55 (1 \pm 0.26) \times 10^{6}$ &  2.26  & $8.29(27.2)\times 10^{-2}$ & \cite{huang:2013geomodel} (global)\tnote{7} \\
\hline
 	\end{tabular}
\begin{tablenotes}
\item[1] The cut-off of Fluka analysis.
\item[2] Average of several theoretical models.
\item[3] Fermi-Dirac spectrum with neutrino temperature of $T_\nu = 3, 5, 8$~MeV for $\nu_e$, $\overline{\nu}_e$, $\nu_x$, respectively.
\item[4] $\nu_x$ is the total contribution from all other neutrinos and antineutrinos.
\item[5] Only the most dominant element is considered. 
\item[6] Combined result from multiple nearby reactors.
\item[7] Global Earth model, incorporates several theoretical models.
\end{tablenotes}
\caption{\label{tab:nucomponents} Neutrino flux components that contribute to the coherent neutrino scattering background in direct detection experiments at the  SNOLab location.~Contributions from solar, atmospheric, diffuse supernovae, reactor as well as geo--neutrinos are shown.}
    \label{tab:nu_flux}
\end{threeparttable}
  \end{center}
\end{table*}

\underline{Solar neutrinos}:~The majority of neutrinos reaching the Earth's surface are produced in nuclear fusion processes in the Sun (for review see \eg ~\cite{Haxton:2012wfz,Bahcall:2004pz}). The flux and energy spectrum of solar neutrinos depends on the specific nuclear reaction chain step that produces them. The vast majority of Sun's energy, nearly 99\%, originates from proton-proton (pp) chain reactions, producing pp, hep, pep,
$^7$Be, $^8$B neutrinos. The remaining $\sim1\%$ of Sun's energy originates from Carbon-Nitrogen-Oxygen (CNO) cycle, producing $^{13}$N, $^{15}$O, $^{17}$F neutrinos. 

For $E_{\nu} \lesssim 20$ MeV energies, solar neutrinos dominate background for direct DM detection experiments. 
In this work we employ solar neutrino fluxes of high metallicity solar model B16-GS98~\cite{Vinyoles:2016djt}, favored by recent Borexino neutrino observation data analysis~\cite{BOREXINO:2022abl}. 
The dominant neutrino background contribution for a range of parameters stems from $^8$B neutrinos.  

\underline{DSNB}:~For neutrino energies around $20~\text{MeV} \lesssim E_{\nu} \lesssim 50$~MeV~DSNB significantly contributes to neutrino background fluxes~\cite{Beacom:2010kk}. 
Around  99\% of the gravitational binding energy is released in the form of neutrinos in core-collapsing supernova, resulting in $\sim10^{58}$ neutrinos emitted yielding about $\sim 10^{53}$~ergs in energy. The DSNB denotes combined contributions of neutrinos originating from all the historic core-collapse supernovae events. The DSNB flux is formed from convolution of the redshift-dependent core-collapse supernovae rate, which is found from initial stellar mass function and star formation rate, as well as emitted core-collapse neutrino spectrum.
Approximating the emitted neutrino spectra as thermal, we consider DSNB neutrinos characterized by temperature $T_{\nu}$ averaged for each species over theoretical models in Ref.~\cite{Horiuchi:2008jz}.
Significant efforts are underway by neutrino experiments, especially Super-Kamiokande~\cite{Super-Kamiokande:2021jaq}, to detect the DSNB in the near future.

\underline{Atmospheric neutrinos}:~For $E_{\nu} \gtrsim 50$ MeV, 
neutrino contributions become dominated by neutrinos produced from cosmic ray collisions with nuclei in the atmosphere~(see \eg~\cite{Gaisser:2002jj} for review). Such interactions produce copious amounts of $\pi^\pm$ and $K^\pm$ mesons that predominantly decay to $\mu^+ + \nu_\mu$ and $\mu^- + \bar{\nu}_\mu$, followed by muon decay to electrons and pairs of neutrinos resulting in a $\nu_\mu$ and $\bar{\nu}_\mu$ flux twice that of $\nu_e$ and $\bar{\nu}_e$.
The atmospheric neutrino flux carries an $\mathcal{O}(30\%)$ theoretical uncertainty, which we take to be 50\% conservatively, due to the uncertainty in the cosmic ray flux and its propagation in the magnetic fields of the Earth and solar system. In addition to this uncertainty, the flux varies significantly though predictably with the position of the detector and the phase in the solar cycle~\cite{PhysRevD.105.043001}.

\underline{Neutrino Oscillations}: As shown in Table~\ref{tab:nu_flux}, solar neutrinos are produced as electron neutrinos.  
Since these neutrinos have a much longer baseline than those in reactor experiments, neutrino oscillations result in a significant fraction of neutrinos reaching the detector with $\mu$- or $\tau$-components, allowing experiments to constrain interactions of second and third generation neutrinos.

Neutrino oscillations in matter are further complicated by the Mikheyev–Smirnov–Wolfenstein (MSW) effects~\cite{Wolfenstein:1977ue}. The matter potential $V_{\rm MSW} = \sqrt{2}G_F n_e \textrm{diag}(1, 0, 0)$, where $n_e$ is the electron number density, modifies the neutrino oscillation Hamiltonian as 
\begin{equation}
    \mathcal{H}_{\rm matter} = E_\nu + \frac{M M^\dagger}{2E_\nu} + V_{\rm MSW}~.
\end{equation}
Vector NSI contribute to the matter potential analogously to MSW effects~\cite{Friedland:2004pp}. Scalar NSI contribute a new mass-like term, which has inverse neutrino energy dependence and hence suppressed at higher neutrino energies~\cite{Ge:2018uhz}. We consider these effects and include in oscillation analysis as discussed in Appendix~\ref{sec:oscilations}. Throughout, we assume normal neutrino mass ordering.

% In the two-flavor neutrino mixing approximation the conversion probability is given by 
% \begin{equation} \label{eq:mixing}
%     P(\nu_i \rightarrow \nu_j) = \sin^2(2\theta_{ij})\sin^2(\pi L/L_{osc})\, ,
% \end{equation}
% where $i,j$ label the neutrino flavor. For $\nu_e-\nu_\mu$ mixing, we consider~\cite{ParticleDataGroup:2022pth}  $\sin^2(\theta_{12})=0.31$, $\Delta m_{12}^2=7.5 \times 10^{-5}{\rm eV}^2$, and $L_{osc} = 2.48{~\rm km} \times E_\nu / \Delta m_{12}^2$ with $E_\nu$ in GeV. Here, $L$ is the baseline distance in km that for solar neutrinos is the Earth-Sun distance. 

\section{Nonstandard Neutrino Interactions} \label{sec:BSM}

There are several formalisms to incorporate general interactions of neutrinos (see \eg~\cite{10.21468/SciPostPhysProc.2.001}). One system of simplified models which has gained popularity generates NSI by introducing new mediators which couple to neutrinos and normal fermions as outlined in Ref.~\cite{Cerdeno:2016sfi}, also employed in e.g. Ref.~\cite{lindner_coherent_2017, Schwemberger:2022fjl}.  categorizes interactions by the Lorentz structure of the mediator such that interactions are defined as scalar, pseudoscalar, vector, axial vector, and tensor (SPVAT).
Specifically, we focus on scalar mediators with masses above $\sim 1$ TeV, which allows us to map our general interactions to effective four-Fermi interactions at low energies which are recognizable in the conventional language of NSI~\cite{10.21468/SciPostPhysProc.2.001}.

In this work, we will consider either flavor independent couplings of a generalized scalar interaction, or what is often called a ``minimal coupling'' scenario, in which all couplings are zero except the one of interest that couples a particular flavor of neutrino with a specific quark. The latter is typically more useful in mapping to UV complete models, and example of which we show in Sec.~\ref{sec:LQ}. For this case although the form of the cross-section is flavor-independent, the fluxes vary as discussed in Sec. \ref{sec:flux}. This means that for a comprehensive treatment of a scalar coupled to multiple flavors one would need to calculate the scattering of each flavor independently and combine the signals. Additionally, we are primarily interested in couplings to first generation quarks as this is where DM experiments are sensitive as we discuss in Sec. \ref{sec:disc_reach}. In Sec.~\ref{sec:neutral} we provide the expressions for scalar NSI and in Sec.~\ref{sec:charged}, we do the same for vector NSI.

\subsection{Scalar NSI} \label{sec:neutral}

The effective Lagrangian for scalar NSI with Dirac neutrinos is given by~\cite{Ge:2018uhz}
\begin{equation} \label{L_eff}
    \mathcal{L}_{sNSI} \supset G_F\sum_{q,\alpha,\beta}\varepsilon_{\alpha\beta}^{qS} \bar{\nu}_\alpha\nu_\beta \bar{q}q\, ,
\end{equation}
where the interaction is a dimension six operator suppressed by two powers of a large mass scale. The indices $q$ run over the quark flavors, while $\alpha,\beta$ run over the lepton flavors. In this work, we will consider flavor-conserving interactions, $\alpha=\beta=\ell$. Scalar NSI cannot be converted to conventional vector-like NSI by Fierz transformations since they arise from a neutral scalar mediator~\cite{PhysRevD.101.095029}. On the other hand, as we will discuss in Sec.~\ref{sec:LQ}, a charged (colored) scalar produces a conventional vector NSI after Fierz transformations.

\begin{figure}
    \centering
    \includegraphics[width=0.6\textwidth]{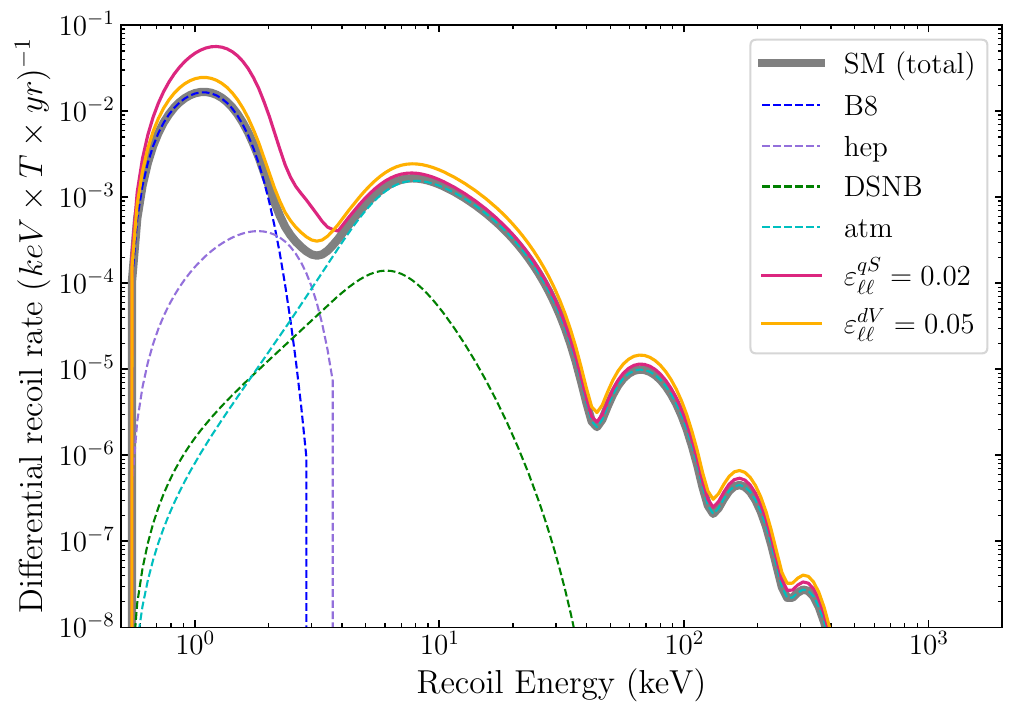}
    \caption{Recoil rates in a xenon detector with a 5 keV threshold for SM interactions of neutrinos (dashed lines), with B8, hep, DSNB, and atmospheric neutrino sources, and the total (thick grey line) as well as the total including a scalar NSI (magenta) for $\varepsilon^{qS}_{\ell\ell} = 0.02$ for all $q$ and $\ell$.}
    \label{fig:recoil_rates}
\end{figure}

The scalar NSI effective vertex gives rise to an additional contribution to the CE$\nu$NS cross-section:

\begin{equation} \label{eq:n_cross}
    \left.\frac{d\sigma}{dE_R}\right\vert_{S} \hspace{0.5mm}  =  \frac{F^2(E_R)G_F^2 \varepsilon_{q\ell}^2 q_S^2 E_R m_N^2}{4\pi E_\nu^2}\, ,
\end{equation}

\noindent where $E_R$ is the energy of the recoiling quark, $m_N$ is the nuclear mass, and $q_S$ is the effective scalar charge of the nucleus. In the flavor/generation independent case, we have $q_S \approx 14A + 1.1Z$~\cite{Cerdeno:2016sfi}.
A sample of the differential recoil rate, both from the SM and NSI is shown in Fig. \ref{fig:recoil_rates}.

\subsection{Vector NSI} \label{sec:charged}

The effective Lagrangian for a vector NSI is

\begin{equation}\label{eq:vec_lag}
    \mathcal{L}_{NSI} = -2\sqrt{2}G_F \sum_{q, \alpha, \beta} \varepsilon^{qP}_{\alpha\beta} \left(\bar{\nu}_\alpha \gamma^\mu P_L \nu_\beta\right) \left(\bar{q}\gamma_\mu P q \right)~,
\end{equation}
where $P=P_L,P_R$ are the projection operators. 
As with the neutral scalar, we assume flavor-conserving NSI, that is, $\alpha=\beta=\ell.$
Similarly, we assume the interaction conserves quark flavor. 
Vector NSIs can be parameterized in terms of vector ($\varepsilon^{qV}_{\ell \ell} = \varepsilon^{qL}_{\ell \ell} + \varepsilon^{qR}_{\ell \ell}$) and axial ($\varepsilon^{qA}_{\ell \ell} = \varepsilon^{qL}_{\ell \ell} - \varepsilon^{qR}_{\ell \ell}$) components.

In the case of CE$\nu$NS, the vector component of the interaction gets enhanced by the size of the nucleus while the axial component is proportional to the nuclear spin. This results in a 1/$A$ suppression of the axial component relative to the vector part. As in \cite{Barranco:2005yy}, we drop the axial component and consider only the vector component of the interaction.
The scattering cross-section for vector NSI is given by
\begin{equation}\label{eq:vec_cross}
    \left.\frac{d\sigma}{dE_R}\right\vert_{NSI} \hspace{0.5mm} = \frac{F^2(E_R)G_F^2 Q_{NSI}^2 m_N}{2\pi E_\nu^2}\left(2E_\nu^2 - m_N E_R\right)~.
\end{equation}
In Eq. \eqref{eq:vec_cross}, $Q_{NSI}$ is the change in the effective nuclear coupling induced by the NSI and given by

\begin{equation}
    Q_{NSI}^2 = \left[ Q_{SM} + \left(2\varepsilon_{\ell \ell}^{uV} + \varepsilon_{\ell \ell}^{dV}\right)Z + \left(\varepsilon_{\ell \ell}^{uV} + 2\varepsilon_{\ell \ell}^{dV} \right) N\right]^2 - Q_{SM}^2
\end{equation}

\noindent for a nucleus with $Z$ protons and $N$ neutrons. $\theta_W$ is the Weinberg angle and $Q_{SM} = \left[\left(1/2 - 2\sin^2(\theta_W)\right)Z - N/2\right]$ is the SM contribution to the neutrino-nuclear coupling. In general, since the $\varepsilon_{\ell\ell}^{qV}$ parameters can be positive or negative, there exists parameter space for a given detector material where the up and down couplings cancel and the NSI does not change the scattering rate. 

In what follows, we will study two cases which we call 
``flavor universal" and ``minimal" NSI, where ``flavor universal" means $\varepsilon_{ee}=\varepsilon_{\mu\mu}=\varepsilon_{\tau\tau}$. In the scalar NSI case, this extends to the quark sector as well, that is $\varepsilon^{uS}_{\ell\ell} = \varepsilon^{dS}_{\ell\ell}$. In the vector NSI case, we note it is impossible for a singlet scalar to have both $\varepsilon^{dV}_{\ell\ell}$ and $\varepsilon^{uV}_{\ell\ell}$ while conserving electric charge, so in the case of vector NSI, ``flavor universal" refers only to lepton flavors. ``Minimal" NSI means $\varepsilon^{qV}_{\ell\ell}\neq 0$ for only one flavor of $\ell$ and $q$.

\section{Case Study: Scalar Leptoquarks} \label{sec:LQ}

We will demonstrate how models of leptoquarks (LQs), which interact with baryons and leptons, provide an example application of our formalism. LQs are theoretically well-motivated and naturally arise in the context of Grand Unification theories such as $SU(5)$~\cite{Georgi:1974sy} and $SU(10)$~\cite{Fritzsch:1974nn,Georgi:1974my}, as well as the Pati-Salam models that unify quark and leptons~\cite{Pati:1974yy}. Further, they can manifest in $R$-parity violating models of supersymmetry~\cite{Hall:1983id,Barbier:2004ez}.

LQs have been recently connected to possible observational anomalies in semi-leptonic $B$ decays~\cite{HFLAV:2019otj,LHCb:2021trn,angelescu_closing_2018,Lee:2021jdr}. In addition, scalar LQs have been posed as solutions to the neutrino mass problem~\cite{Dorsner:2017wwn, PhysRevD.LQ3} (see Appendix~\ref{sec:nu_mass} for an example). If the LQ couples to muons, it provides a contribution to the anomalous magnetic moment at one loop order and may be part of the explanation for the anomalous magnetic moment of the muon $(g-2)_\mu$~\cite{zhang_radiative_2021,dorsner_muon_2020, FileviezPerez:2021lkq,gherardi_low-energy_2021}.

\begin{figure}
\includegraphics[width=1.0\textwidth]{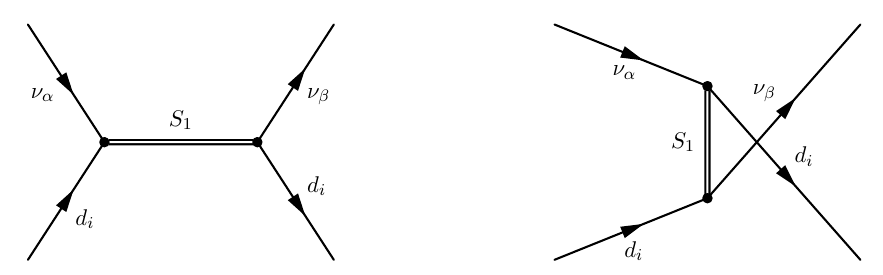}
     \caption{Neutrino-quark scattering mediated by the $S_1$ scalar leptoquark.}
     \label{fig:LQ_scattering}
\end{figure}

Here, we consider LQs in the large coupling regime, which are of interest for observations and provide a heavy scalar that couples to both quarks and neutrinos~\cite{Schmaltz:2018nls, DORSNER20161}. 
To map the LQ models to the NSIs discussed above, we note
that because the mediator carries lepton and quark quantum numbers, the interactions proceed via $s$ and $u$-channel diagrams rather than the $t$-channel of uncharged scalar interactions (see Fig.~\ref{fig:LQ_scattering}). Once the massive fields are integrated out, a Fierz transformation leads to an effective four-fermion vertex identical to the vector NSI Sec.~\ref{sec:charged}. We derive the vector NSI parameters in Sec.~\ref{sec:couplings}.

\subsection{The \texorpdfstring{$S_1$}{S1} leptoquark} \label{sec:s_1LQ}

We will focus on scalar LQs as the simplest realization of neutrino NSIs. 
As a concrete realization, we consider the $S_1 = (\bar{3}, 1, 1/3)$ leptoquark, where parenthesis represent transformation under SM $\left( SU(3), SU(2), U(1)\right)$ symmetries. The additional relevant Lagrangian terms due to $S_1$ are~\cite{AristizabalSierra:2007nf, Babu:2019mfe, Babu:2020hun}: 
\begin{equation}\label{eq:S_1}
    \mathcal{L}_{S_1} \supset  -m_{S_1}^2|S_1|^2 + (\lambda_L^{i\ell} S_1 \bar{Q}^c_i \epsilon L_\ell + \lambda_R^{i\ell} S_1^* \bar{u}^i_R \bar{\ell}_R + h.c.)\, ,
\end{equation}
where $Q_i=(u_L,d_L)_i$ and $L_\ell=(\nu_{\ell L},\ell_L)$ are the left-handed quark and lepton doublets respectively, $\ell$ is the lepton flavor, $\epsilon$ is the Levi-Civita tensor, and $i$ is the quark generation index. From expansion of the $SU(2)$-doublets in Eq.~\eqref{eq:S_1}, we note that the same Yukawa interaction that couples charged leptons to up-type quarks will couple neutrinos to down-type quarks. Thus, by constraining neutrino scattering, we also place constraints on the charged lepton $S_1$ couplings. Throughout, we do not consider possible effects of diquark terms, which could lead to nucleon decay. Such terms can be readily restricted by additional symmetries, however detailed model implementation is beyond the scope of the present work.
  
As stressed in Refs.~\cite{Diaz:2017lit,Schmaltz:2018nls}, without a model-independent reason dictating a flavor pattern of LQ couplings all nine quark-lepton generation pairings are possible. Strong constraints exist on the first generation couplings, $\lambda^{qe}$, particularly from atomic parity violating experiments~\cite{Roberts:2014bka,Babu:2019mfe,Qweak:2018tjf,Schmaltz:2018nls}. Here, we shall focus on LQ couplings to the $\mu$ and $\tau$ flavored leptons. 
For simplicity, we will also assume the LQ couples only to one-generation (sometimes called the ``minimal leptoquark"~\cite{Schmaltz:2018nls}), either $\mu$ or $\tau$ leptons.\footnote{The flavor-changing constraints in this case are much weaker and couplings to both $\mu$ and $\tau$ may be simultaneously allowed as long as one coupling is marginally smaller than the other~\cite{Zhang:2021dgl}.}

The interactions of $S_1$ LQ can lead to distinct phenomenological consequences.
Through the coupling to $\mu$-sector the $S_1$ LQ can directly contribute to the magnetic moment of the muon. 
For $m_{S_1} \gg m_t$ the resulting contribution to $a_\mu = (g-2)_\mu/2$ is~\cite{Bauer_muon}
\begin{equation}\label{eq:g-2}
    a^{S_1}_\mu = \sum_i \frac{m_\mu m_{u_i}}{4\pi^2m_{S_1}^2}\left(\ln{m_{S_1}^2/m_{u_i}^2} - 7/4\right)\Re(\lambda_{R}^{*i2}\lambda_L^{i2}) - \frac{m_\mu^2}{32\pi^2m_{S_1}^2}\left(\sum_i |\lambda_R^{i2}|^2 + \sum_i |\lambda_L^{i2}|^2\right)\, ,
\end{equation}
which is dominated by the term proportional to the up-type quark masses. In a LQ model where the charm and top quarks have the same coupling as the up-quark, we can estimate a favored region in the LQ mass vs. up-type Yukawa plane that alleviates the $(g-2)_\mu$ anomaly. This region is indicated with a grey band in Fig.~\ref{fig:LQ_reach}. In the case that the coupling to charm and top quarks vanishes or $\lambda_R \ll \lambda_L$, the second term dominates and exacerbates the anomaly. 

Since we are interested in coupling to protons and neutrons, we restrict ourselves to couplings to first generation of SM quarks. We are considering neutrino-nuclear scattering in or near the elastic regime, so the parton distribution functions (PDFs) are dominated by the first generation quarks.\footnote{This neglects the ``sea quarks'' which provide trace amounts of anti-quarks of the first generation as well as (even fewer) strange quarks and anti-quarks. In principle, these sea quarks provide an effective charge allowing detection of NSIs coupled to strange quarks or, as in the case of the $\tilde{R}_2$ LQ (defined in Appendix~\ref{sec:nu_mass}), NSIs which only couple neutrinos to anti-quarks (or quarks to anti-neutrinos), though the PDF suppression is such that neutrino scattering is not a competitive means to set bounds.} Thus, we consider scalar $S_1$ LQ with couplings $\lambda^{d\ell}$, where $\ell=\mu,\tau$.

With these assumptions and for an $\mathcal{O}$(TeV) LQ, we can integrate our the LQ field, perform a Fierz transformation, and map the resulting dimension six operator back to the NSI parameters in eq. \eqref{eq:vec_lag} as
\begin{equation}
    \varepsilon^{dV}_{\ell\ell,S_1} = \frac{-|\lambda^{d\ell}_L|^2}{4\sqrt{2} G_F m_{S_1}^2}\, .
\end{equation}

We provide details on this calculation in Appendix \ref{sec:couplings}. Note the interactions only involve down-type quarks, which is accounted for by a change in the effective charge $Q_{S_1}$. 
\begin{equation}
    Q_{S_1} = Z \left(\frac{1}{2} - 2\sin^2(\theta_W) + \varepsilon^{dV}_{\ell\ell,S_1}\right) + N \left(-\frac{1}{2} + 2 \varepsilon^{dV}_{\ell\ell,S_1}\right)\, ,
\end{equation}
Lastly, we are considering the ``minimal LQ" model in which the $S_1$ couples to a single flavor of neutrinos, such that the neutrino-nuclear scattering signal only involves a single neutrino-flavor. Note that this is contrast to SM CE$\nu$NS, which involve all flavors.
 
\section{Discovery Reach of Dark Matter Detectors}\label{sec:disc_reach}

We are interested in calculating the potential of a DM direct detection experiment to uncover signals of neutrino NSIs, with the SM CE$\nu$NS as our effective background. In the following analysis, we will present our results in terms of the general NSI parameters $\varepsilon^{qS}_{\ell\ell}$ and in terms of the LQ model parameters, $\lambda_{S_1}^{d\ell}, m_{S_1}$, from the previous section. For generality, we will consider that we are dominated by statistics and do not focus on making predictions in the context of more detector-specific systematic backgrounds, such as radioactive decay in or near the detector, or any other instrumental noise. 
Neutrino oscillation effects are present for both scalar and vector NSIs, though as we discuss in Appendix~\ref{sec:oscilations}, these effects are minimal for the NSI parameters we consider and are not included in the following analysis. 
We also do not consider the presence of a DM signal in this analysis, although Ref.~\cite{Bertuzzo:2017tuf} studied the effect of NSIs on the neutrino fog. 

\subsection{Statistical analysis} \label{sec:stat}

Since we have reduced the NSI cross-section to one free parameter, $\varepsilon^{qS}_{\ell\ell}$, we use a likelihood ratio method to project the significance of simulated data. This follows the method used for the magnetic moment calculation in~Ref.~\cite{Schwemberger:2022fjl}, which was also dependent on only one parameter. We take as our likelihood function
\begin{equation}\label{likelihood}
    \mathcal{L}(\varepsilon, \Vec{\phi}) = \frac{e^{-\Sigma_{j=1}^{n_\nu}(\eta_{\varepsilon,j} + \eta_{SM,j})}}{N!} \times \prod_{j=1}^{n_\nu}\mathcal{L}(\phi_j) \times \prod_{i=1}^{N}\left(\sum_{j=1}^{n_\nu}\left( \eta_{\varepsilon, j} f_{\varepsilon,j}(E_i) + \eta_j f_{SM,j}(E_i) \right)\right)\, ,
\end{equation}
where we have dropped the labels on $\varepsilon=\varepsilon^{qS}_{\ell\ell}$ for simplicity, $n_\nu$ label the neutrino fluxes and $N$ labels the energy bins. Here, $\Vec{\phi}$ are nuisance parameters normalizing the neutrino flux channels. They are allowed to fluctuate weighted by Gaussian distribution, 
\begin{equation}
\mathcal{L}(\phi_j)=\frac{1}{\sigma_j\sqrt{2\pi}}e^{-\frac{1}{2}\left(\frac{\phi_j-\phi_{0,j}}{\phi_j}\right)^2}    \, ,
\end{equation}
where the width of their uncertainties, $\sigma_j$ is given in Table~\ref{tab:nu_flux}. The functions $f_{\varepsilon, j}(E_i)$ and $f_{SM, j}(E_i)$ are the distribution functions (\emph{i.e.} the normalized rates) for the number of events in the $i$-th energy bin from the $j$-th neutrino flux from NSI ($\varepsilon$) or SM scattering. $\eta_{\varepsilon, j}$ and $\eta_{SM, j}$ are the number of events predicted from the NSI or SM in each channel ($j$) based on the nuisance parameters. The latter depends only on the normalization of the neutrino flux components while the former depends on both the flux normalization and the NSI parameter ($\varepsilon$). The product over $i$ runs over the number of events in the simulated data.

For a particular data set, we maximize the likelihood with $\varepsilon=0$ as our null hypothesis and $|\varepsilon|>0$ for NSI signals.  If the null likelihood is greater than the signal likelihood, 
we assign the significance $\sigma=0$. If $\mathcal{L}(\varepsilon \neq 0)/\mathcal{L}(\varepsilon = 0) \equiv \lambda > 1$, we have a test statistic $t = 2\log(\lambda)$ and significance $\sigma = \sqrt{t}$. This analysis is run on 600 of sets of pseudo-data generated by Poisson fluctuating the expected rates for a given $\varepsilon^{qS}_{\ell\ell}$ and exposure. The average of these data sets is then used for the projections in Sec.~\ref{sec:projections}.

\subsection{Results} \label{sec:projections}

\begin{figure}[H]
    \centering 
    \centering
    \begin{subfigure}[t]{0.49\textwidth}
        \centering
        \includegraphics[width=\textwidth]{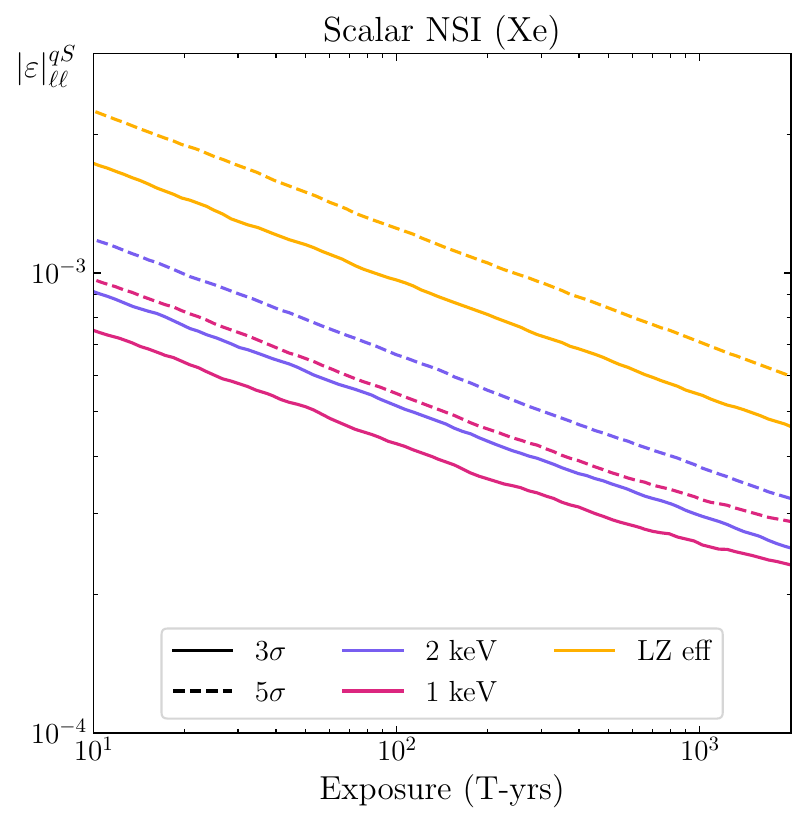}
    \end{subfigure}%
    ~ 
    \begin{subfigure}[t]{0.49\textwidth}
        \centering
        \includegraphics[width=\textwidth]{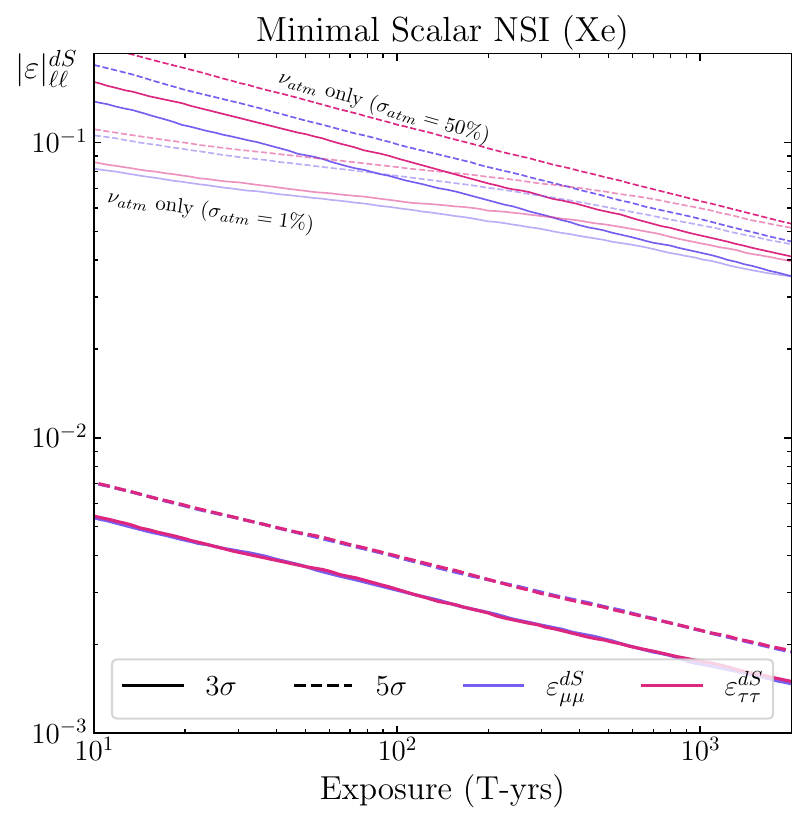}
    \end{subfigure}
    \begin{subfigure}[t]{0.49\textwidth}
        \centering
        \includegraphics[width=\textwidth]{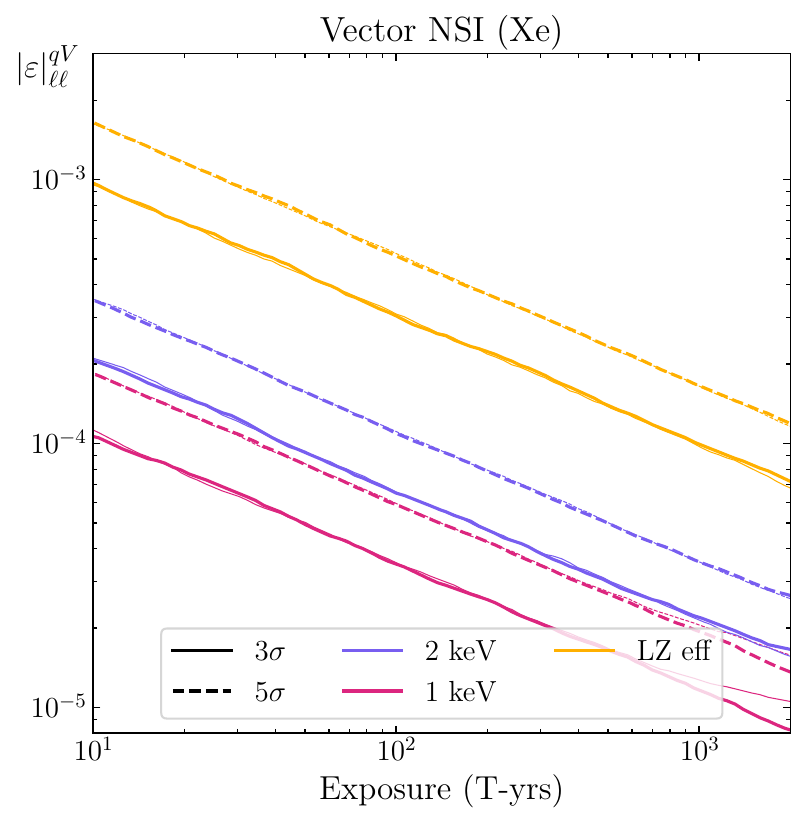}
    \end{subfigure}%
    ~ 
    \begin{subfigure}[t]{0.49\textwidth}
        \centering
        \includegraphics[width=\textwidth]{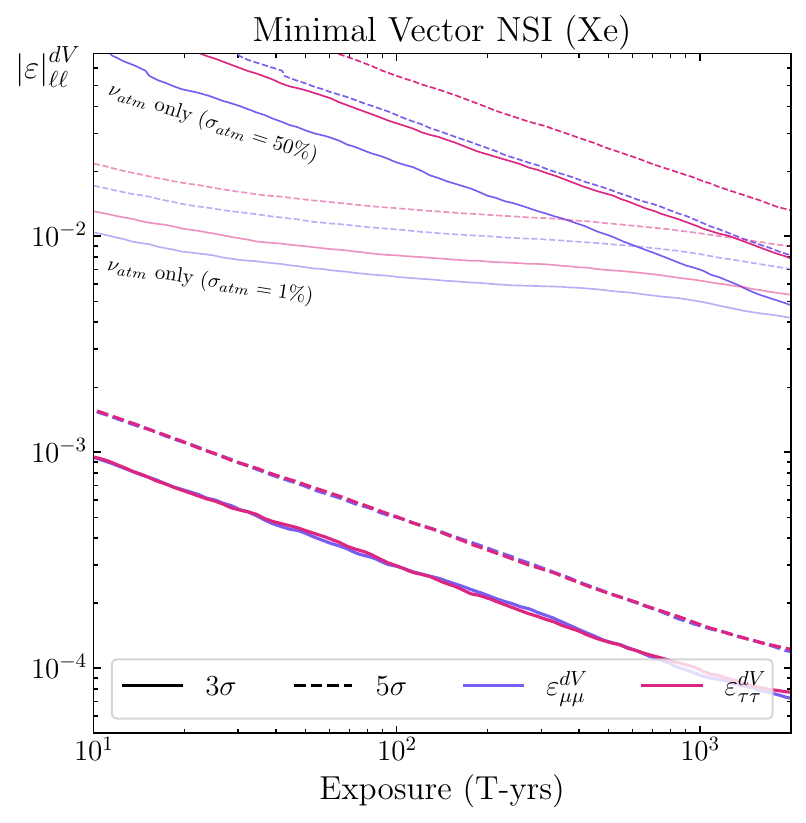}
    \end{subfigure}
    \caption{{\bf Left:} $3\sigma$ (solid) and $5\sigma$ (dashed) discovery reach of a xenon detector for an NSI with universal couplings ($\varepsilon^{dV}_{ee}=\varepsilon^{dV}_{\mu\mu}=\varepsilon^{dV}_{\tau\tau}$) as a function of detector exposure, for a 1 keV (magenta) and 2 keV (purple) detector threshold as defined in Sec.~\ref{sec:rates}. Also shown is the reach using the LZ efficiency and threshold (yellow). As expected, a lower threshold improves the discovery reach. 
    {\bf Right:} discovery reach of a xenon detector to NSIs coupling only the $d$ quark to either $\nu_\mu$ or $\nu_\tau$, assuming a 2 keV threshold. Thin lines indicate the sensitivity based only on atmospheric neutrinos. Fainter lines show the improvement to be gained from more precise measurement of the atmospheric neutrino flux with reduced uncertainties. Upper panels show results for scalar NSI while lower panels are vector NSI results. In the vector NSI panels, we present results for both $\varepsilon<0$ ($\varepsilon>0$) with the thick (thin) lines, and show that they are similar. The lower right panel maps to the $S_1$ LQ parameters in Fig.~\ref{fig:LQ_reach}.}
    \label{fig:thresholds}
\end{figure}

In this section, we present the results of the likelihood procedure described in the previous section. We show the 3$\sigma$ and 5$\sigma$ contours for $\varepsilon$ in the exposure-$\varepsilon$ plane in Fig.~\ref{fig:thresholds}, for a flavor universal scalar ($\varepsilon^{qS}_{\ell\ell}$, upper left) and a minimal scalar ($\varepsilon^{dS}_{\ell\ell}$, upper right) while the equivalent contours for vector NSI are shown in the lower panels. We find that the discovery regions of the experiments we consider are within the range of values allowed by IceCube ($-0.041 < \varepsilon^{qV}_{\ell\ell} < 0.042$)~\cite{PhysRevD.104.072006} and Super-Kamiokande ($-0.049 < \varepsilon^{qV}_{\ell\ell} < 0.049$)~\cite{PhysRevD.84.113008}. For the flavor universal scalar, we show the effects of varying the detector energy threshold. The yellow curves show the reach using the efficiency from the LZ experiment~\cite{LZ:2022ufs}. Improving the detector efficiency can lead to a significantly stronger reach, and we see that lowering the detector threshold from $E_{\rm th}=2$ keV to 1 keV improves the reach on $\varepsilon$ by about a factor of few. Since $P_{e\to\tau} > P_{e\to\mu}$ when averaged over the solar neutrino spectrum, there will be $\sim 3$ $\nu_\tau$ for every 2 $\nu_\mu$, thus we expect $\tau$ constraints to be stronger by a factor $\sim \sqrt{3/2}$. This is confirmed by the results in Fig.~\ref{fig:thresholds} where we also show the results of an analysis of only atmospheric neutrinos which are useful to constrain interactions with anti-neutrinos such as the $\tilde{R}_2$ LQ (see Appendix~\ref{sec:nu_mass}). We further display how the results will improve as atmospheric neutrino flux uncertainty is further reduced, as far down as optimistic $\mathcal{O}(1\%)$-level.

\begin{figure}[h]
    \centering
    \hspace{-1.3cm}
    \includegraphics[width=0.95\textwidth]{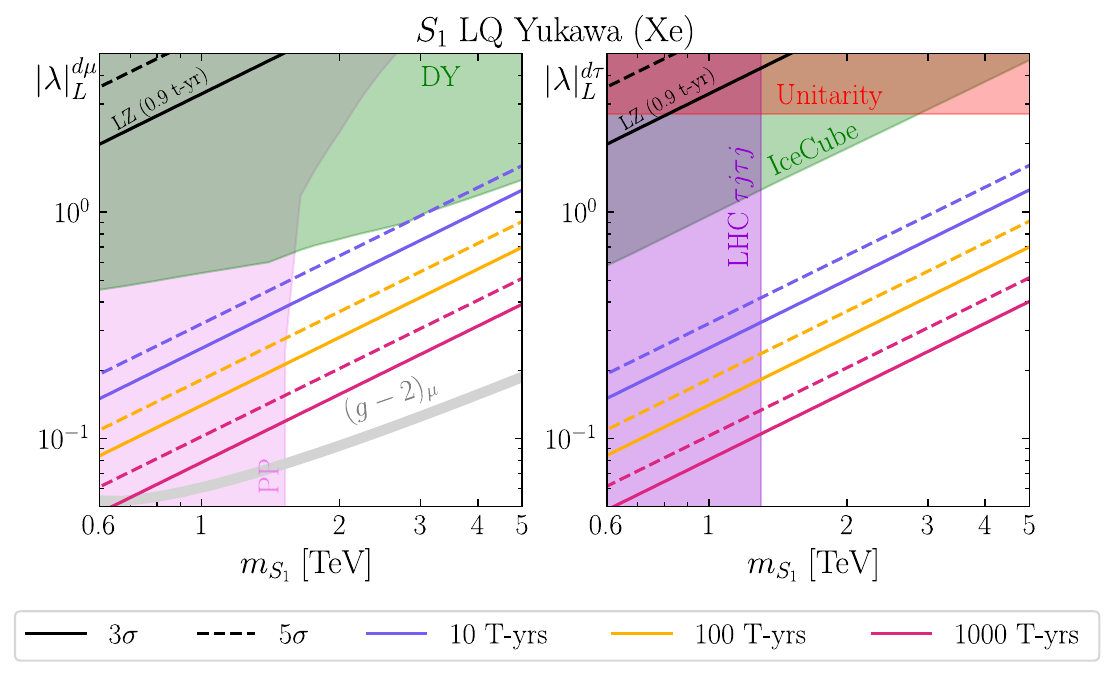}
    \caption{Projected constraints on the $\mu-d$ (left) and $\tau-d$ (right) Yukawas for a selections of exposures, mapped from sensitivity to a $\varepsilon^{dV}_{\ell\ell}$ minimal NSI in a xenon DM detector with a 2 keV threshold. Dashed (solid) lines represent 5$\sigma$ and 3$\sigma$ sensitivity reach. We set the first limits on LQs from preliminary LZ data with 0.9 ton-year exposure \cite{LZ_eff}. In the $d-\mu$ figure (left), the thick grey line indicates the preferred values to alleviate the $(g-2)_\mu$ anomaly for quark generation and chirality independent couplings. Pair production (PP) and Drell-Yan (DY) bounds on the muon coupling are from 36 fb$^{-1}$ data from the LHC \cite{PhysRevD.99.032014, sirunyan_search_2018}. In the $d-\tau$ figure (right), we show the LHC bounds on the tau coupling~\cite{atlascollaboration2023search} which constrain $m_{LQ} > 1.3$ TeV. IceCube results are from resonance searches in atmospheric neutrino data \cite{Babu:2022fje}. At the highest masses, the LQ Yukawa coupling is only constrained by perturbative unitarity~\cite{allwicher_perturbative_2021}.
    }
    \label{fig:LQ_reach}
\end{figure}
We also present our results in terms of the scalar LQ $S_1$ parameters, $\lambda^{d\ell}$ and $m_{S_1}$, in Fig.~\ref{fig:LQ_reach} for a few select detector exposures. Our work, for the first time, shows the constraints from the LZ experiment in black. In addition, we show constraints from the LHC searches looking for LQ pair production (PP) and Drell-Yan (DY) production using 36 fb$^{-1}$ of data~\cite{PhysRevD.99.032014, sirunyan_search_2018}. As seen in Fig.~\ref{fig:LQ_reach}, the DM direct detection constraints are competitive with the LHC constraints for the $\lambda^{d\mu}$, but do not access the parameter region relevant for $(g-2)_\mu$ shown in gray. The DM direct detection constraints are significantly stronger than the LHC constraints for $\lambda^{d\tau}$, where the mass and short lifetime of the charged $\tau$ weaken the LHC constraints~\cite{Schmaltz:2018nls}, and the IceCube constraints coming from atmospheric neutrinos~\cite{PhysRevD.104.072006}. For both scenarios, there are constraints from perturbative unitarity~\cite{allwicher_perturbative_2021}, which provide the strongest constraints on $\lambda^{d\tau}$ for $m_{S_1}\gtrsim 3$ TeV.

In addition to xenon, other DM experiments can also have enhanced sensitivity to CE$\nu$NS and related new physics. In general, the effective charge of the nucleus grows as the square of its mass, so for a given detector target mass, larger nuclei are expected to result in more events even though there are fewer target nuclei. This competes with the kinematic effects whereby higher intensity but lower energy neutrinos are often pushed below threshold in detectors with heavy nuclei. 

\begin{figure}
    \centering
    \begin{subfigure}[t]{0.45\textwidth}
        \centering
        \includegraphics[width=\textwidth]{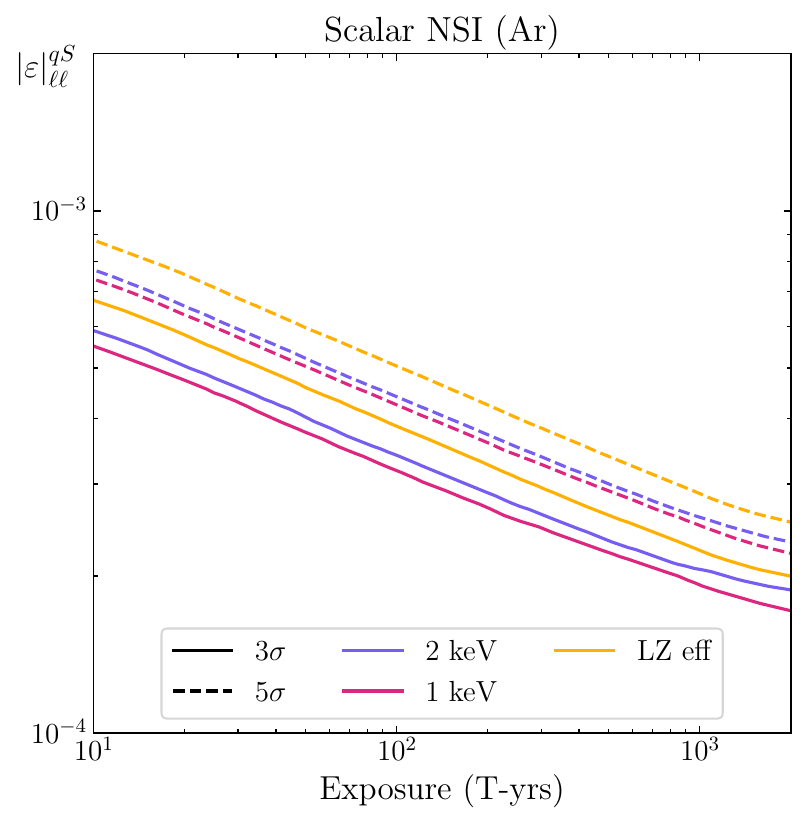}
    \end{subfigure}%
    ~ 
    \begin{subfigure}[t]{0.45\textwidth}
        \centering
        \includegraphics[width=\textwidth]{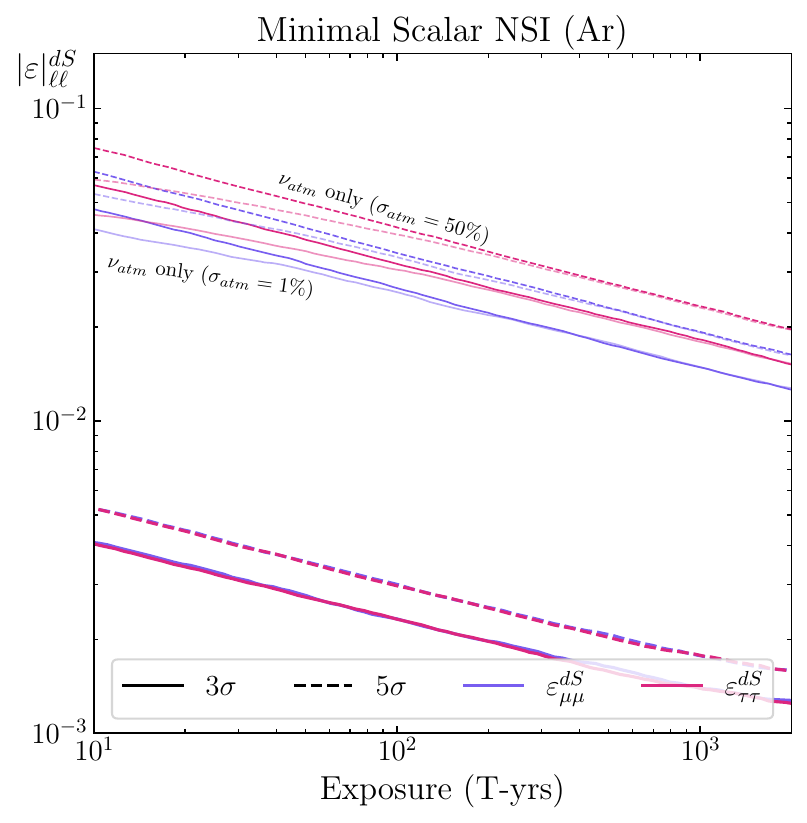}
    \end{subfigure}
    \begin{subfigure}[t]{0.45\textwidth}
        \centering
        \includegraphics[width=\textwidth]{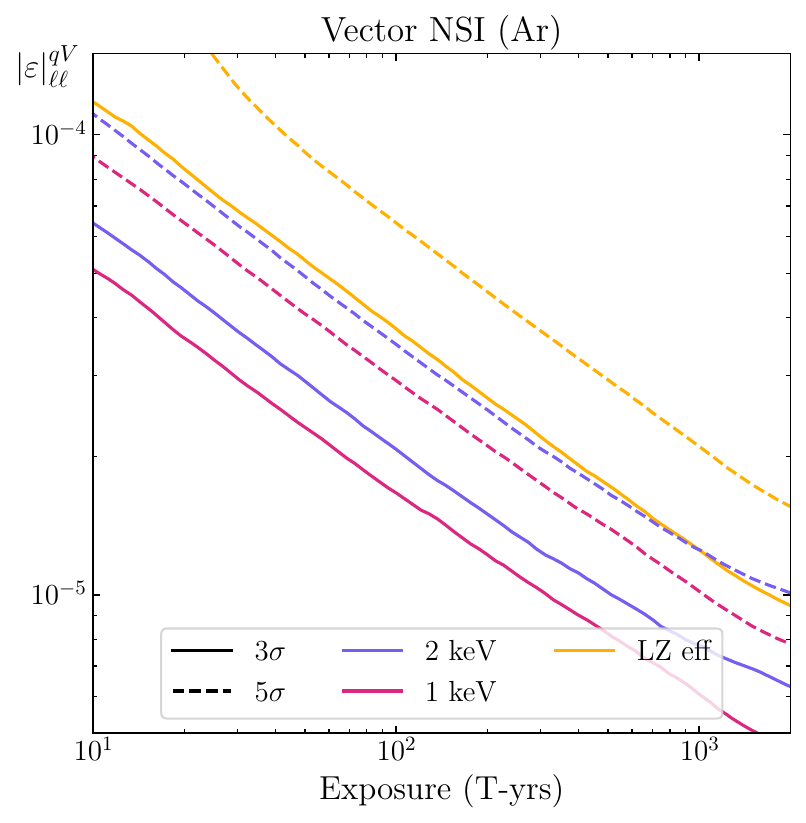}
    \end{subfigure}%
    ~ 
    \begin{subfigure}[t]{0.45\textwidth}
        \centering
        \includegraphics[width=\textwidth]{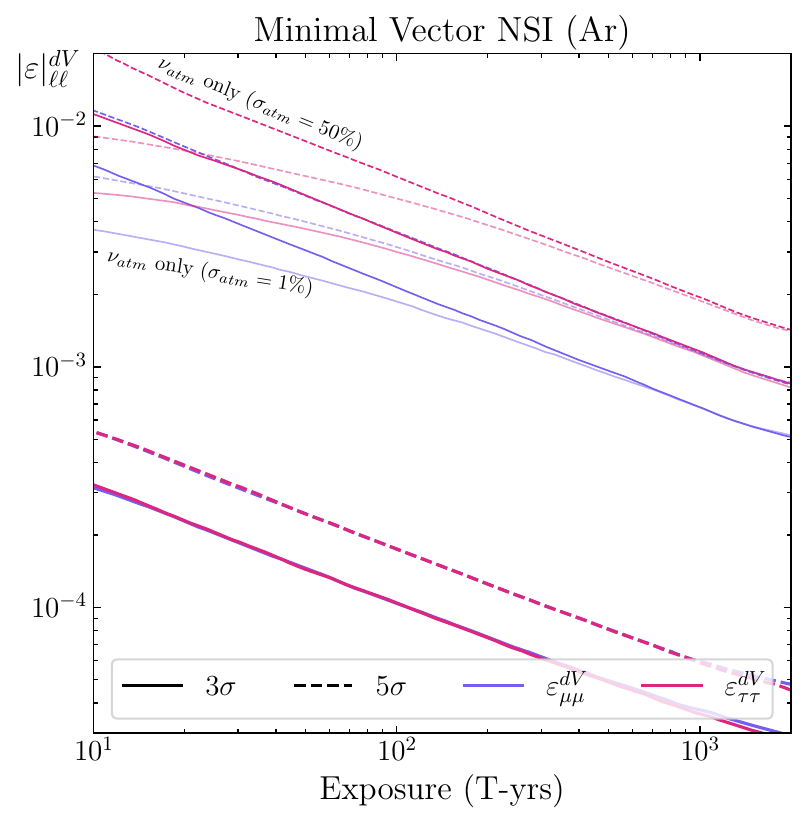}
    \end{subfigure}
    \caption{{\bf Left:} $3\sigma$ (solid) and $5\sigma$ (dashed) discovery reach of an argon detector for a scalar (top) or vector (bottom) NSI with universal couplings as a function of detector exposure, for a 1 keV (magenta) and 2 keV (purple) detector threshold as defined in Sec.~\ref{sec:rates}. As expected, a lower threshold improves the discovery reach. Due to its larger target mass, kinematics disfavors lead detectors despite the increased scattering rate. This is also responsible for the more dramatic improvement with detector threshold in lead.
    {\bf Right:} discovery reach of an argon detector to scalar (top) or vector (bottom) NSIs coupling only the $d$ quark to either $\nu_\mu$ or $\nu_\tau$, assuming a 2 keV threshold. Thin lines indicate the sensitivity based only on atmospheric neutrinos. Fainter lines show the improvement to be gained from more precise measurement of the atmospheric neutrino flux.}
    \label{fig:Ar_NSI_reach}
\end{figure}

\begin{figure}
    \centering
    \begin{subfigure}[t]{0.45\textwidth}
        \centering
        \includegraphics[width=\textwidth]{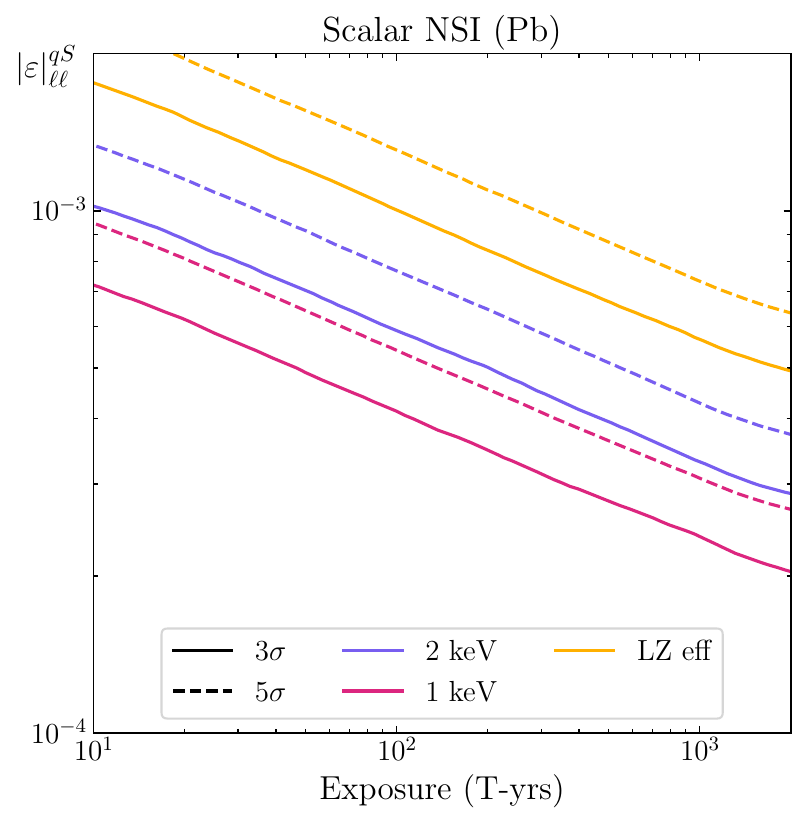}
    \end{subfigure}%
    ~ 
    \begin{subfigure}[t]{0.45\textwidth}
        \centering
        \includegraphics[width=\textwidth]{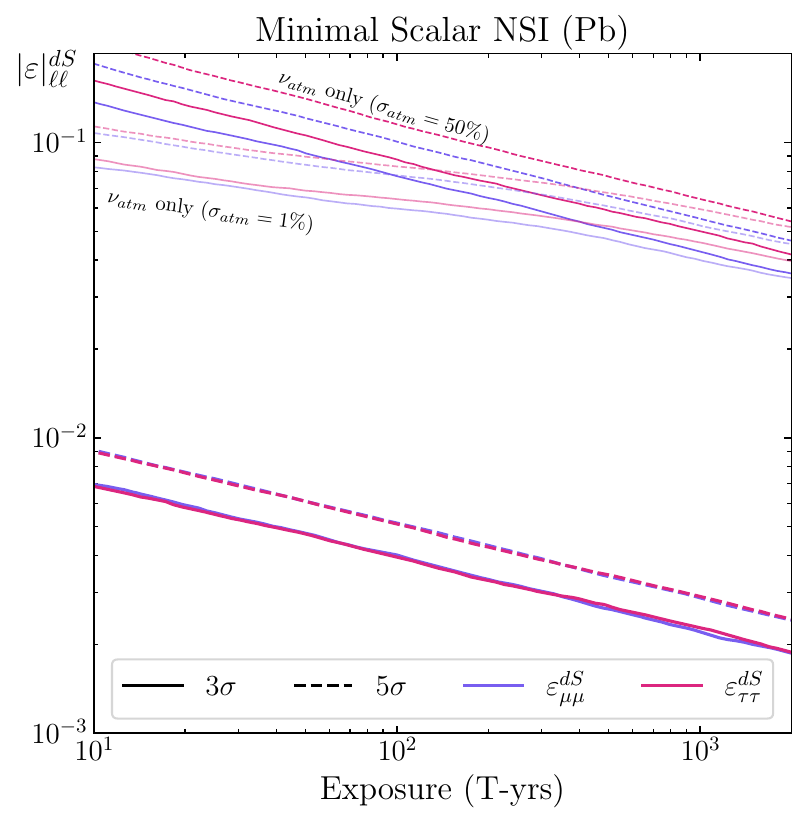}
    \end{subfigure}
    \begin{subfigure}[t]{0.45\textwidth}
        \centering
        \includegraphics[width=\textwidth]{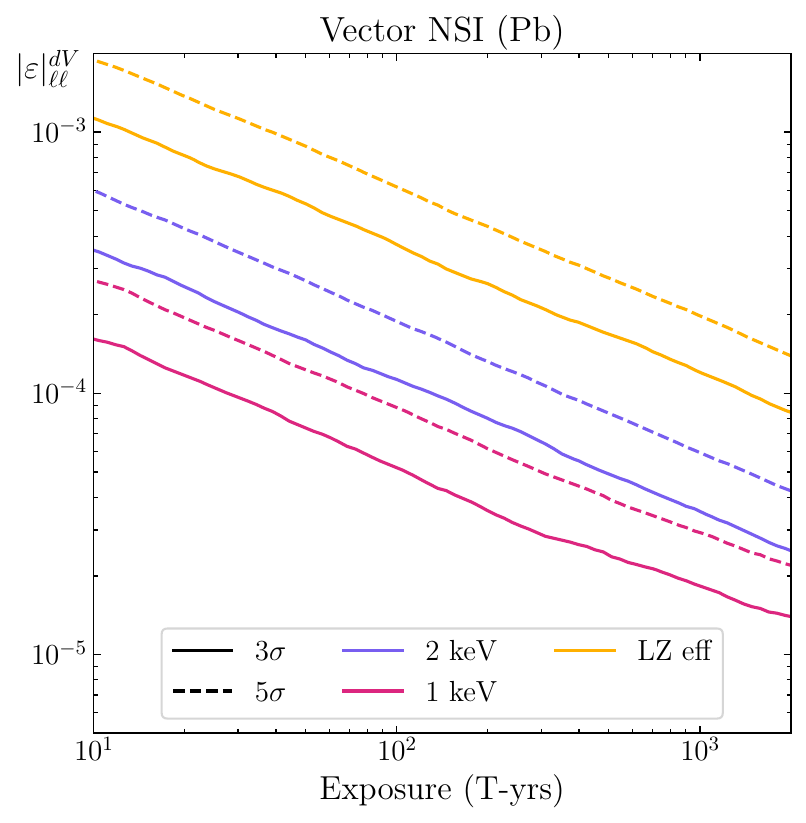}
    \end{subfigure}%
    ~ 
    \begin{subfigure}[t]{0.45\textwidth}
        \centering
        \includegraphics[width=\textwidth]{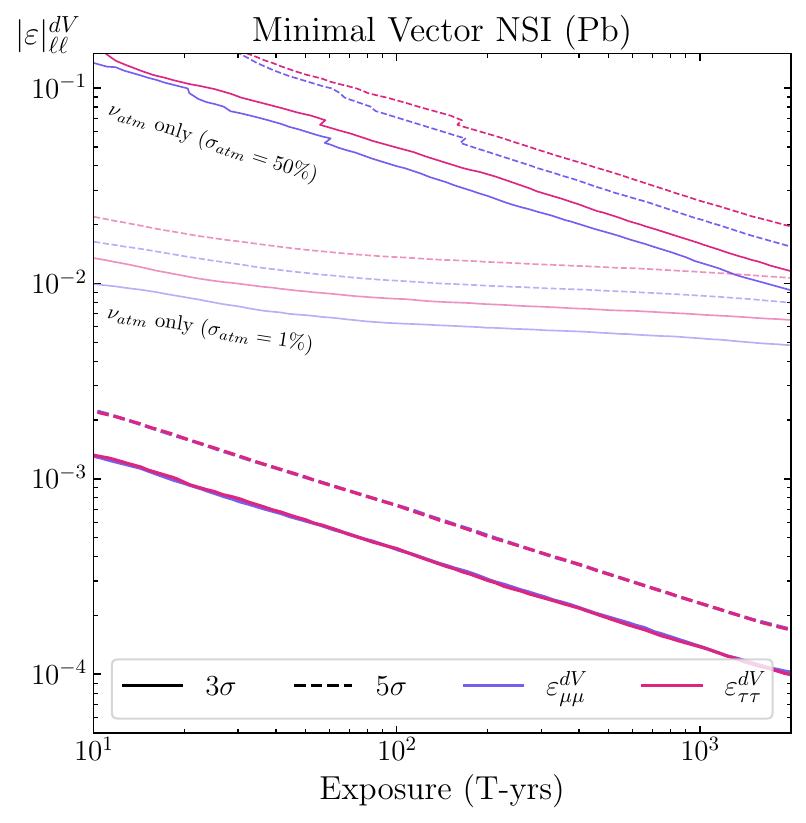}
    \end{subfigure}
    \caption{Same as Fig.~\ref{fig:Ar_NSI_reach}, but for a lead detector. Note that due to its heavier nucleus, lead loses more sensitivity than argon or xenon at higher thresholds, and gains more from higher energy neutrinos such as the atmospheric flux.}
    \label{fig:Pb_NSI_reach}
\end{figure}

In figures~\ref{fig:Ar_NSI_reach} and \ref{fig:Pb_NSI_reach} we display the 3$\sigma$ and 5$\sigma$ discovery reach contours for the NSI parameter $\varepsilon$ in the exposure-$\varepsilon$ plane for argon and lead-based detectors. For estimating the sensitivity to scalar NSIs, while the details depend on the specific experimental configuration, we assume for simplicity the same threshold and efficiency parameters as for xenon-based detector.
Note that argon does not see much improvement as the threshold falls below 5 keV (the differences seen above are statistical noise). Argon has a relatively light nucleus and the peak induced by boron-8 is at higher energy, so lowering the threshold does not increase the number events as efficiently until the oxygen-16 neutrinos have an effect (below $\sim$ 0.5 keV). For the heavier lead nucleus, recoils induced by boron-8 neutrinos are partially cut by the threshold, so improving the threshold makes a much larger number of recoils accessible. The lower panels of figures~\ref{fig:Ar_NSI_reach} and \ref{fig:Pb_NSI_reach} show the sensitivity to minimal couplings assuming the 2 keV threshold and efficiency.

\begin{figure}
    \centering
    \includegraphics[width=\linewidth]{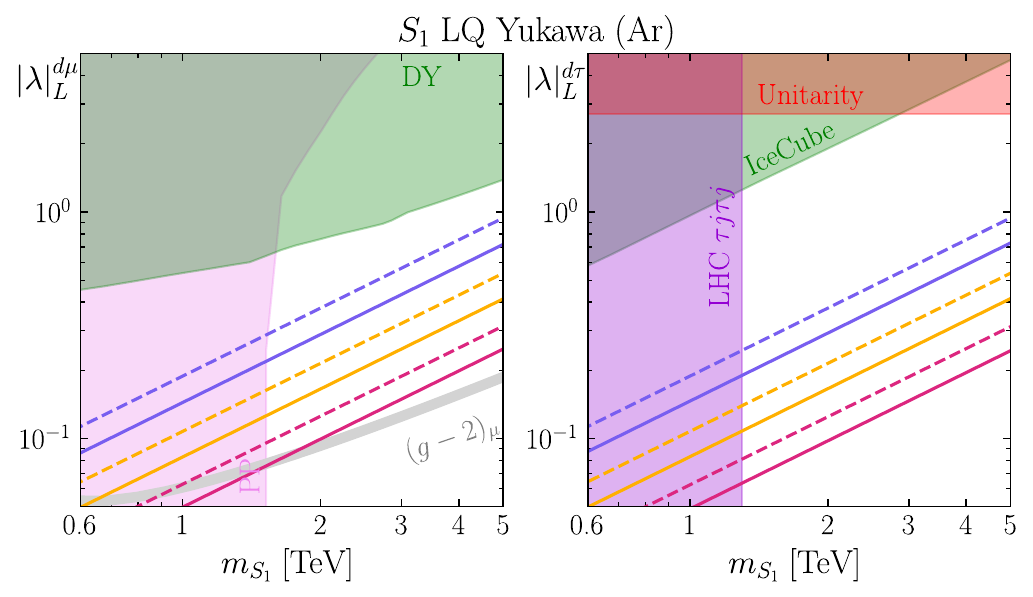}
    \includegraphics[width=\linewidth]{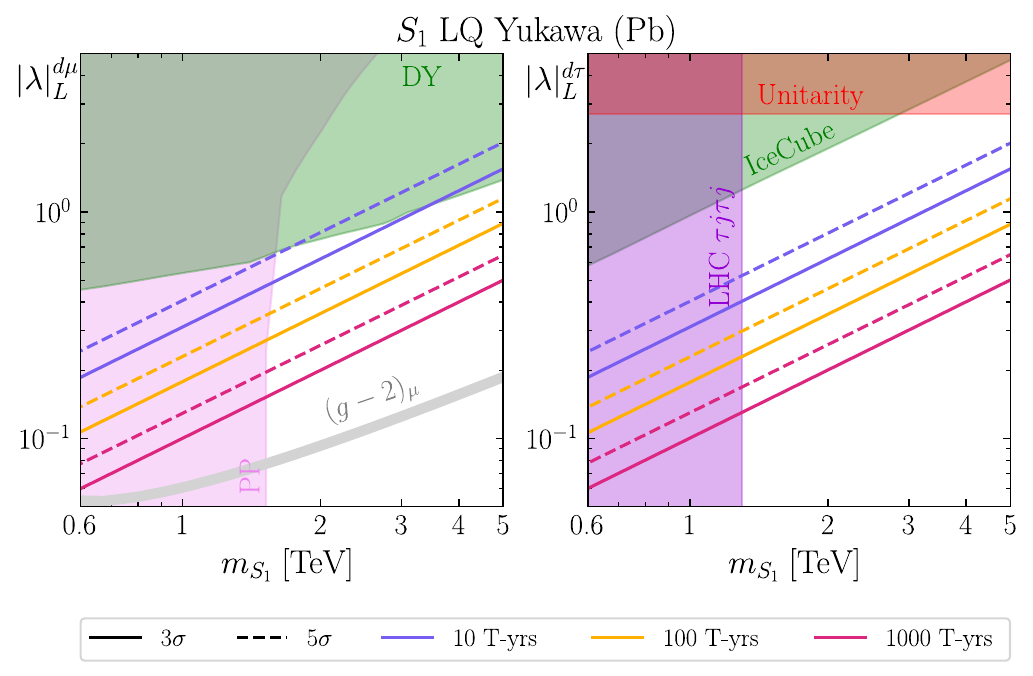}
    \caption{Same as Fig. \ref{fig:LQ_reach}, but for argon (top) and lead (bottom).}
    \label{fig:scalar_Y_Pb}
\end{figure}

In Fig.~\ref{fig:scalar_Y_Pb}, we translate our results into sensitivity projections for scalar LQ $S_1$ parameters, $\lambda^{q\ell}_L$ and $m_{S_1}$, as before. We find that a detector constructed from either material, argon or lead, and reaching multi-ton-year exposure can be sensitive and improve existing bounds on the $S_1$ LQ couplings.

\section{Conclusions}
Direct DM detection experiments constitute a central pillar in the search for DM and will continue to increase their exposures and lower thresholds and backgrounds. As we demonstrate, this will allow such experiments to constrain or reveal new physics in the space of neutrino interactions, complementary to conventional neutrino telescopes. We set new limits on LQs associated with scalar neutrino NSI using latest data of direct DM detection experiments, in particular LZ. We find that near-future DM detectors can probe parameter space of LQs that is out of reach of the LHC. We discuss how upcoming measurements improving uncertainties for atmospheric neutrino could probe LQ and NSI parameter space. The studied LQs are of particular interest for interpreting the observations of $(g-2)_\mu$ and explaining neutrino masses. Our analysis can be extended to variety of other models and similar methodology could also be applied to vector-like LQs, which may be associated with measurements and anomalies in flavor physics.
Such a realization of massive scalar mediated NSIs provides a new benchmark for testing new physics beyond the SM at planned near-future experiments.

\acknowledgments
\addcontentsline{toc}{section}{Acknowledgments}

We thank Pouya Asadi, Kaladi Babu, Bhupal Dev, Bhaskar Dutta, Motoi Endo, Danny Marfatia for helpful discussions. T.S. and T.-T.Y. were supported in part by NSF CAREER grant PHY-1944826. T.-T.Y. also thanks the CCPP at NYU for hospitality and support. V.T.
was supported in part by the World Premier International Research Center Initiative (WPI), MEXT, Japan and also by JSPS KAKENHI grant No. 23K13109.

\appendix

\section{Model extension and neutrino mass}\label{sec:nu_mass} 
Following Ref.~\cite{Babu:2019mfe}, we discuss a simple model extension of the $S_1$ LQ with a second scalar LQ, $\tilde{R}_2 = (3, 2, 1/6)$, that is an $SU(2)$ doublet $(R^{+2/3}, R^{-1/3})$, which can readily result in Majorana neutrino mass generation at one-loop level. The considered LQs can arise in a variety of theoretical frameworks, such as models of $R$-parity violating supersymmetry~\cite{Hall:1983id,Barbier:2004ez}.
The additional Lagrangian terms due to $\tilde{R}_2$ are
\begin{equation}\label{R_2}
    \mathcal{L}_{\tilde{R}_2} \supset (\tilde\lambda^{i\ell}\bar{d}^i_R\tilde{R}_2\varepsilon L_\ell + \mu H^{\dagger} \tilde{R}_2 S_1^{\ast} + h.c.) -m_{\tilde{R}_2}^2|\tilde{R}_2|^2 ~,
\end{equation}
where the $\mu$-term mixes $S_1$ and $\tilde{R}_2$ LQs and $H = (1,2,1/2)$ is the SM Higgs doublet.

\begin{figure}
\centering
\includegraphics[angle=0,width=0.5\textwidth]{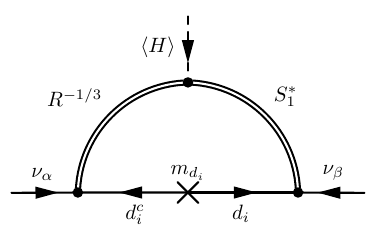}
    \caption{One-loop neutrino mass generation in the $S_1 + \tilde{R}_2$ LQ model. }
    \label{fig:nu_mass}
\end{figure}

At loop-level, $S_1$ and $\tilde{R}_2$ LQs generate neutrino mass, as displayed in Fig.~\ref{fig:nu_mass}.
Expanding the $SU(2)$ doublets, the necessary terms to generate neutrino NSI and masses are

\begin{equation}
    \mathcal{L}_{S_1, \tilde{R}_2} \supset \tilde\lambda^{i\ell} \left(\nu_\ell \bar{d}_i R^{-1/3} - \ell \bar{d}_i R^{+2/3}\right) + \lambda_L^{i\ell} \left(\nu_\ell d_i - \ell \bar{u}^c_i\right) S_1 - \mu R^{-1/3} H^0 S_1^* + h.c.
\end{equation}
The mixing term implies the existence of a new LQ mass basis with eigenvalues
\begin{equation}
    m_{1,2}^2 = \frac{1}{2}\left(m_{R^{+2/3}}^2 + m_{S_1}^2 \mp \sqrt{(m_{R^{-1/3}}^2 - m_{S_1}^2)^2 + 4\mu^2v^2}\right)
\end{equation}
and a mixing angle 
\begin{equation}
    \tan(2\theta) = \frac{-\sqrt{2}\mu v}{m_{S_1}^2 - m_{R^{-1/3}}^2}~,
\end{equation}
where $v$ is the Higgs vev and $m_{R^{-1/3},S_1}$ include the bare masses and the mass induced by spontaneous symmetry breaking. The neutrino mass matrix is given by

\begin{equation} \label{eq:m_nu}
    M_\nu = \frac{3\sin(2\theta)}{32\pi^2}\log\left(\frac{m_1^2}{m_2^2}\right) (\tilde\lambda M_d \lambda_L^T + \lambda_L M_d \tilde\lambda^T)
\end{equation}
where $M_d$ is a diagonal matrix of down-type quark masses. In the case of a LQ coupled only to down-type quarks and leptons, Eq.~\eqref{eq:m_nu} simplifies to 
\begin{equation}
    (M_\nu)^{\alpha \beta} = \frac{3\sin(2\theta)}{32\pi^2}\log\left(\frac{m_1^2}{m_2^2}\right) m_d \times (\lambda_L^{d\alpha} \tilde\lambda^{d\beta} + \tilde\lambda^{d\alpha} \lambda_L^{d\beta})\, ,
\end{equation}
and similarly for LQs coupled to only strange or bottom quarks with $m_d \to m_{s,b}$. In the specific cases we study, only one element of $\lambda^{d\alpha}_L$ is non-zero implying only the generation of a single neutrino mass. However, the $\tilde{R}_2$ coupling $\tilde\lambda^{d\alpha}$ may contribute a second element to the neutrino mass matrix, or as noted in \ref{sec:s_1LQ}, it is possible for both $\lambda^{d\mu}_L$ and $\lambda^{d\tau}_L$ to be non-zero.

We note that with a multitude of parameters determining the neutrino mass above, there is degeneracy in LQ coupling and mass parameter space. From cosmological observations, the sum of neutrino masses given by the eigenvalues of $M_\nu$ is constrained to $\sum m_\nu < 0.13$ eV by cosmic microwave background radiation (CMB) data from Planck and the Dark Energy Survey (DES)~\cite{PhysRevD.107.023531}. 

\section{From leptoquarks to NSI}\label{sec:couplings}

To map the $S_1$ LQ to the scalar NSI parameters, we need to find the expressions for $\varepsilon^{dV}_{\ell\ell}$ and $q_S$ in terms of the LQ parameters. Expanding the $SU(2)_L$ doublets in eq. \eqref{eq:S_1} and explicitly writing the hermitian conjugate, we have
\begin{equation}
    \mathcal{L}_{S_1}\supset -m_{S_1}^2 S_1^*S_1 + \lambda_L^{1\ell} \left( \bar u^c P_L \ell - \bar d^c P_L \nu_\ell \right) S_1 + \lambda_L^{*1\ell} \left( \bar u^c P_L \ell - \bar d^c P_L \nu_\ell \right)^\dagger S_1^*
\end{equation}
Where we've specified $q=1$ for the $u$ and $d$ quarks.
Since we are considering massive scalars of $\mathcal{O}$(TeV), we can integrate out the LQ using the equations of motion
\begin{equation}
\begin{split}
    \frac{\partial \mathcal{L}_{S_1}}{\partial S^*} &= -m_{S_1}^2 S_1 + \lambda_L^{*1\ell} \left( \bar u^c P_L \ell - \bar d^c P_L \nu_\ell \right)^\dagger = 0 \\
    \frac{\partial \mathcal{L}_{S_1}}{\partial S} &= -m_{S_1}^2 S_1^* + \lambda_L^{1\ell} \left( \bar u^c P_L \ell - \bar d^c P_L \nu_\ell \right) = 0
\end{split}
\end{equation}
and find
\begin{equation}
    \mathcal{L}_{S_1}\supset \frac{|\lambda_L^{1\ell}|^2}{m_{S_1}^2}\left[ (\bar u^c P_L \ell)(\bar u^c P_L \ell)^\dagger - (\bar u^c P_L \ell)(\bar d^c P_L \nu_\ell)^\dagger - (\bar d^c P_L \nu_\ell)(\bar u^c P_L \ell)^\dagger + (\bar d^c P_L \nu_\ell)(\bar d^c P_L \nu_\ell)^\dagger \right]
\end{equation}
We now focus on the last term which introduces an effective four-Fermi interaction between $d$ quarks and neutrinos.
\begin{equation}
    \mathcal{L}_{S_1}\supset \frac{|\lambda_L^{1\ell}|^2}{m_{S_1}^2} (\bar d^c P_L \nu_\ell) (\nu_\ell^\dagger P_L (\bar d^c)^\dagger) = \frac{|\lambda_L^{1\ell}|^2}{m_{S_1}^2} (\bar d^c P_L \nu_\ell) (\bar \nu_\ell P_R d^c)
\end{equation}
After a Fierz transformation~\cite{bischer_loop-induced_2018}, this becomes
\begin{equation}
    \mathcal{L}_{S_1}\supset -\frac{|\lambda_L^{1\ell}|^2}{2m_{S_1}^2} (\bar d^c \gamma_\mu P_R d^c) (\bar \nu_\ell \gamma_\mu P_L \nu_\ell)
\end{equation}
Now we evaluate the charge conjugation of the quarks and find
\begin{equation}
    \mathcal{L}_{S_1}\supset \frac{|\lambda_L^{1\ell}|^2}{2m_{S_1}^2} (\bar d \gamma_\mu P_L d) (\bar \nu_\ell \gamma_\mu P_L \nu_\ell)
\end{equation}
Finally, noting that we get no RH coupling and $\varepsilon^{qV}_{\ell \ell} = \varepsilon^{qL}_{\ell \ell} + \varepsilon^{qR}_{\ell \ell}$ we find that the $S_1$ LQ maps to Eq.~\eqref{eq:vec_lag} with 
\begin{equation}
    \varepsilon^{dV}_{\ell\ell,S_1} = \frac{-|\lambda^{d\ell}_L|^2}{4\sqrt{2} G_F m_{S_1}^2}\, .
\end{equation}
We note that the NSI parameter resulting from an $S_1$ LQ is negative.

\section{Neutrino oscillation effects}\label{sec:oscilations}

In Sec. \ref{sec:flux} we briefly discussed the effects of neutrino oscillations. Here, we expand the discussion and consider neutrino oscillations with and without NSI effects in neutrino propagation through the Earth or Sun. To analyze neutrino propagation through Earth we use the NuCraft code~\cite{Wallraff:2014qka}, which is based on the PREM Earth reference model\footnote{NuCraft was built for IceCube and thus considers a detector 2~km deep near the south pole. Slight deviations are noted with results for different geomagnetic locations including Kamioka in Japan and Sudbury in Canada}~\cite{Dziewonski:1981xy}, and include modifications for vector neutrino NSI~\cite{Friedland:2004pp, Gonzalez-Garcia:2013usa, Coloma:2017egw} and scalar neutrino NSI~\cite{Ge:2018uhz}. For solar neutrinos, full in-medium effects and NSI are already built into the SNuDD module~\cite{Amaral:2023tbs} that we use to calculate the flavor dependence of the solar neutrinos when they reach Earth. In all cases we assume normal neutrino mass ordering.

In the case of vector NSI, the contributions to the matter potential result in the effective Hamiltonian
\begin{equation}
    {\cal H_{\rm vNSI,matter}}\simeq E_\nu+\frac{MM^\dagger}{2E_\nu}\pm(V_{\rm SI}+V_{\rm NSI})\, ,
\end{equation}
where the neutrino mass matrix $M$, SM matter potential $V_{\rm SI}$, and the vector NSI matter potential contributions $V_{\rm NSI}$ are 3$\times$3 matrices. The matter potential becomes significant only when the neutrino energy $E_\nu$ or matter density are sizable $2E_\nu V\gtrsim \Delta m_{ij}^2$~\cite{Ge:2018uhz}.
 % This is to be expected as current experimental limits on non-flavor-violating NSI from oscillations of atmospheric neutrinos are $|\varepsilon_{\ell\ell}|\sim 0.04$~\cite{PhysRevD.104.072006}.
Flavor conserving NSI in the $\mu - \tau$ sector can be parameterized by a single parameter $\varepsilon_{\tau\tau} - \varepsilon_{\mu\mu}$ where $\varepsilon_{\ell\ell} = \sum_f Y_f \varepsilon_{\ell\ell}^{fV}$ is the sum of couplings to each fermion weighted by the fermion number density $Y_f$ (relative to $Y_e$). In this parameterization, current experimental limits are $|\varepsilon_{\ell\ell}|\lesssim 0.04$~\cite{PhysRevD.104.072006} in $\mu - \tau$ sector. Constraints on flavor violating NSI from oscillations are much stronger since they have a greater effect on oscillation probabilities, but we do not consider these NSI in this work.

\begin{figure}
    \centering
    \includegraphics[width=0.7\linewidth]{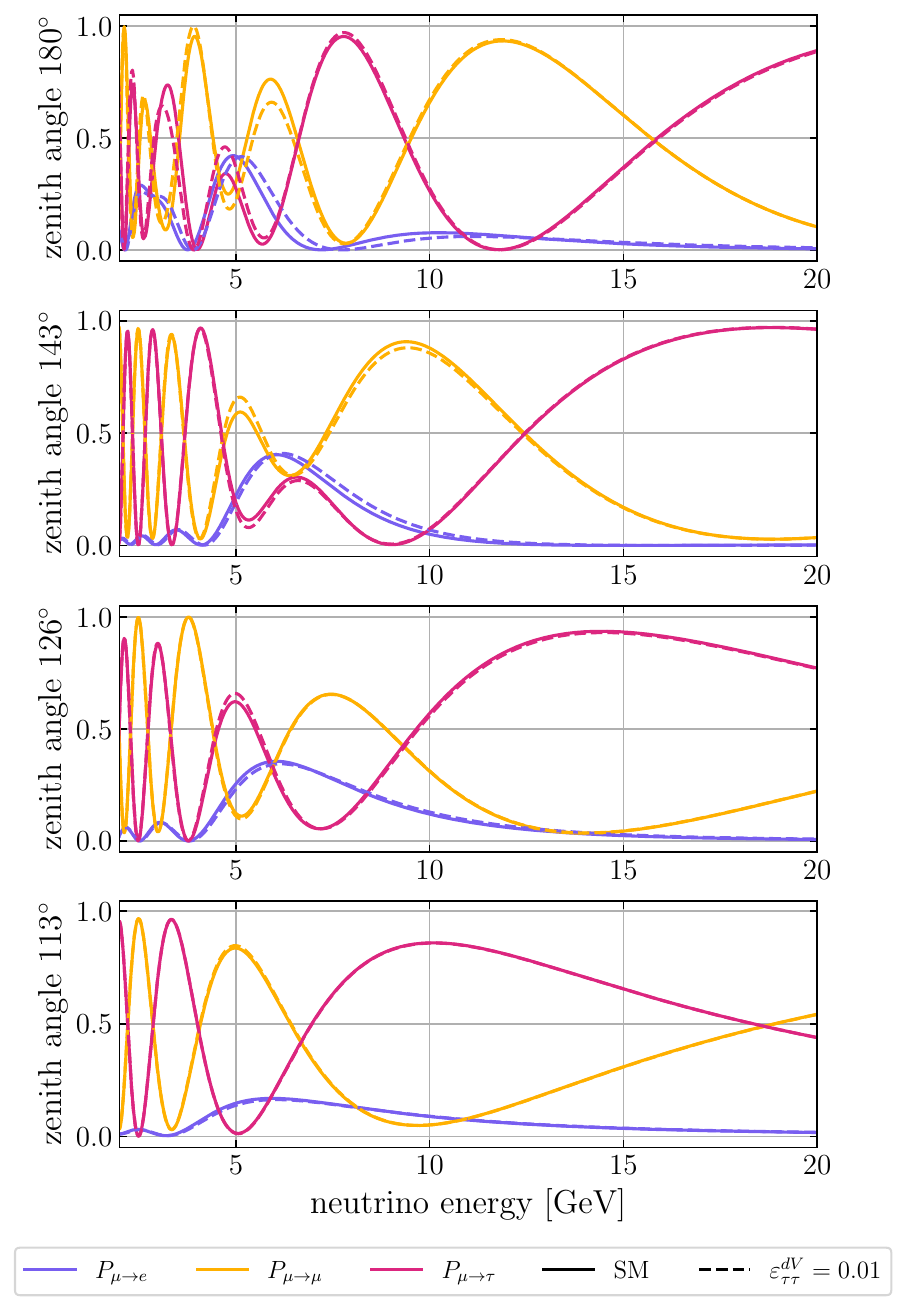}
    \caption{Oscillation conversion probabilities of atmospheric neutrinos, calculated using the NuCraft code, with (dashed line) and without (solid line) vector NSI effects for distinct directions with respect to zenith.}
    \label{fig:atm_osc}
\end{figure}

In Fig.~\ref{fig:atm_osc} we display the oscillation probabilities for a characteristic vector NSI values we consider 
$|\varepsilon^{qV}_{\tau\tau}|\sim 10^{-2}$ along with the SM medium effects inside the Earth. For the SM effects, we find reasonable agreement with~\cite{Super-Kamiokande:2017yvm}. Note that effects of a coupling to up or down quarks are similar, since the NSI potential scales with $Y_u$ and $Y_d$ respectively. From Fig.~\ref{fig:atm_osc} we observe that at particular propagation directions with respect to zenith\footnote{Straight down propagation is at $180^\circ$.} and neutrino energies vector NSI $\varepsilon\sim 0.01$ can modify flavor oscillation by $\mathcal{O}(10\%)$. Here we do not consider detectors with directional sensitivity. Hence, integrating the atmospheric flux contributions over the field-of-view and including recoil kinematics and detector resolution further reduces the significance of the vector NSI on the signal in dark matter experiments.
Further detailed study and consideration of detectors with directional sensitivity could leverage these effects to improve sensitivities to vector NSI, analysis of which we leave for future work.

\begin{figure}
    \centering
    \includegraphics[width=\linewidth]{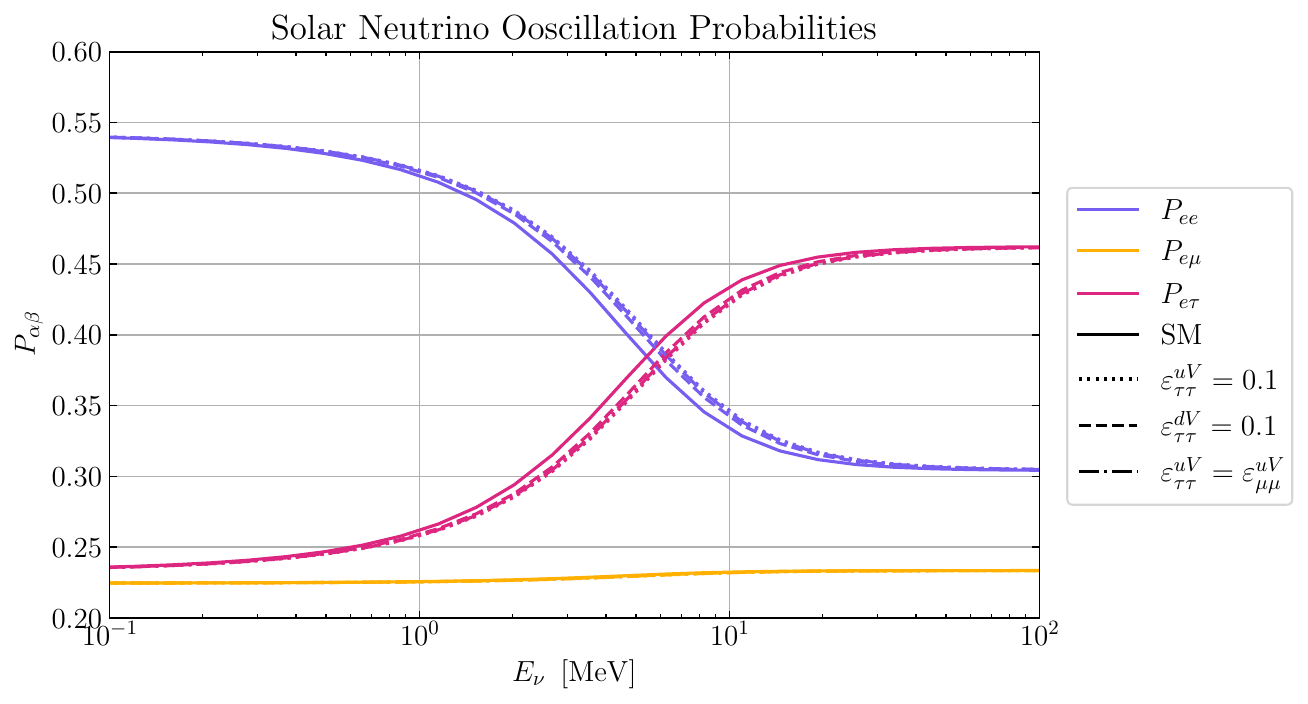}
    \caption{Oscillation conversion probabilities of solar neutrinos, as calculated with the SNuDD module, for SM (solid line) and including vector NSI coupling of tau neutrinos to up quarks (dotted line), down quarks (dashed line), or both (dot-dashed line).}
    \label{fig:sol_osc}
\end{figure}

The vector NSI effects are suppressed at lower neutrino energies and thus the NSI parameters we consider have negligible effect on solar neutrinos despite the increase in medium density. In Fig.~\ref{fig:sol_osc} we display oscillation probabilities for solar neutrinos as calculated using SNuDD package considering NSI of $\varepsilon=0.1$, which is significantly larger than the parameters we consider in this work. Solar neutrino oscillations in vacuum are also considered, though are subdominant to the medium effects. We assume a constant Earth-Sun distance of $1.5\times 10^{8}$~km. Qualitatively, we have found agreement of our considerations with discussion in Ref.~\cite{Maltoni:2015kca}.

In contrast to vector NSI, scalar NSI induce an effective correction to the neutrino mass matrix, which scales with the matter density and is suppressed by the neutrino energy. This leads to a neutrino oscillation probability that is sensitive to the matter density variations along the propagation baseline~\cite{Ge:2018uhz}. 
In the flavor basis, $M_{eff} = M + M_{\rm NSI}$ where $M$ is the standard neutrino mass matrix and $M_{\rm NSI} = \sum_q n_q G_F \varepsilon^{qS}_{\ell\ell}$ where $\varepsilon^{qS}_{\ell\ell}$ is the scalar NSI parameter in Eq.~\eqref{L_eff} and $n_q$ is the quark number density. We focus on flavor conserving NSIs, so $M_{\rm NSI}$ is diagonal in the flavor basis\footnote{In general, $\varepsilon^{qS}_{\alpha\beta}$ are also possible.}. Transforming to the mass basis and rotating the phasing matrix into the NSI term, $M_{eff}$ is given in Eq.~\eqref{eq:m_eff} where $D_\nu = \textrm{diag}(m_1, m_2, m_3)$\footnote{$\textrm{diag}(x, y, z)$ is defined as the $3\times3$ diagonal matrix with elements $x, y, z$ on the diagonal.} is the diagonal neutrino mass matrix, $\mathcal{U}$ is the PMNS matrix, and $P$ is the diagonal rephasing matrix
\begin{equation}\label{eq:m_eff}
    M_{eff} = \mathcal{U}D_\nu\mathcal{U}^\dagger + P^\dagger M_{\rm NSI} P~.
\end{equation}
Following the notation of Ref.~\cite{Ge:2018uhz}, we take $\sqrt{|\Delta m^2_{31}|}$ as a characteristic mass scale and parameterize the NSI mass contribution as
\begin{equation}\label{eq:del_M}
    P^\dagger M_{\rm NSI} P = \delta M = \sqrt{|\Delta m^2_{31}|} (\eta)_{i,j}
\end{equation}
where $(\eta)_{i,j}$ are matrix elements in the mass basis.
For neutrinos propagating in matter with scalar NSI contributions, the effective Hamiltonian is 
\begin{equation}\label{eq:H_nsi}
    \mathcal{H}_{{\rm sNSI}, 
\rm matter} = E_\nu + \frac{M_{eff}M_{eff}^\dagger}{2E_\nu} + V_{\rm SI}~,
\end{equation}
where $V_{\rm SI}$ is the SM matter potential from charged-current interactions. For antineutrinos, $V_{\rm SI}$ changes sign.
At the energies and couplings we consider in the analysis, SM vector interactions contribute significantly more strongly to neutrino oscillations than scalar NSI, such that the latter do not have appreciable effects in dark matter experiments.

\clearpage
\bibliography{refs}
\addcontentsline{toc}{section}{Bibliography}
\bibliographystyle{JHEP}

\end{document}